\documentclass[aps,prx,twocolumn,amsmath,nofootinbib,amssymb,superscriptaddress,10pt,floatfix]{revtex4-1}
\usepackage{graphicx}
\usepackage{ulem}   
\usepackage{xcolor}
\usepackage[colorlinks, linkcolor=blue, citecolor=blue]{hyperref}
\renewcommand{\vec}[1]{\boldsymbol{#1}}

\def \Dw {{\cal{D}}(\omega)}

\def \x {{\vec x}}

\def \D{\Delta}

\def \beq {\begin{eqnarray}}
\def \eeq {\end{eqnarray}}
\def \tn {\textnormal}

\newcommand{\DpRP}{\ensuremath{\delta p}}
\newcommand{\pRP}{\ensuremath{p}}

\begin{document}
\title{Jamming and Unusual Charge Density Fluctuations of Strange Metals}
\author{Stephen J. Thornton}
\affiliation{Department of Physics, Cornell University, Ithaca NY 14853, U.S.A.}
\author{Danilo B. Liarte}
\affiliation{Department of Physics, Cornell University, Ithaca NY 14853, U.S.A.}
\affiliation{ICTP South American Institute for Fundamental Research, S\~ao Paulo, SP, Brazil}
\affiliation{Institute of Theoretical Physics, S\~ao Paulo State University, S\~ao Paulo, SP, Brazil}
\author{Peter Abbamonte}
\affiliation{Department of Physics, University of Illinois at Urbana-Champaign, Urbana IL 61801, U.S.A.}
\author{James P. Sethna}
\author{Debanjan Chowdhury}
\email{debanjanchowdhury@cornell.edu}
\affiliation{Department of Physics, Cornell University, Ithaca NY 14853, U.S.A.}
\date{\today}
\begin{abstract}
The strange metallic regime across a number of high-temperature superconducting materials presents numerous challenges to the classic theory of Fermi liquid metals. Recent measurements of the dynamical charge response of strange metals, including optimally doped cuprates, have revealed a broad, featureless continuum of excitations, extending over much of the Brillouin zone. The collective density oscillations of this strange metal decay into the continuum in a manner that is at odds with the expectations of Fermi liquid theory. Inspired by these observations, we investigate the phenomenology of bosonic collective modes and the particle-hole excitations in a class of strange metals by making an analogy to the phonons of classical lattices falling apart across an unconventional  jamming-like transition associated with the onset of rigidity. By making comparisons to the experimentally measured dynamical response functions, we reproduce many of the qualitative features using the above framework. We conjecture that the dynamics of electronic charge density over an intermediate range of energy scales in a class of strongly correlated metals can be at the brink of a jamming-like transition.
\end{abstract}
\maketitle

\begin{center}
{\bf INTRODUCTION}
\end{center}

A hallmark of numerous interacting phases of quantum matter are their long-lived collective excitations (such as phonons, magnons, and skyrmions). Microscopically, these collective modes require a coherent motion of the constituent particles in the system. While such modes often have a long lifetime at low energies, they are prone to decay once they encounter the multi-particle continuum at high energies. Even in weakly interacting metals, there are two kinds of long-lived excitations --- the plasmon, which represents a collective (longitudinal) density fluctuation, and single-electron like quasiparticle excitations near the Fermi surface. The plasmon eventually decays at large enough momentum and frequency (i.e. for $\omega>\omega_\star(q)$) into the multi-particle continuum due to purely kinematic reasons. Within Landau's original formulation of Fermi liquid  (FL) theory for electrically neutral fermions (e.g. as in liquid Helium-3) \cite{pines}, the zero-sound mode is associated with a collective oscillation of the entire Fermi surface and has properties that are qualitatively similar to a longitudinal acoustic phonon. The sound mode gets renormalized into the plasmon mode in the presence of Coulomb interactions. It is natural to consider the fate of collective modes and their possibly unconventional decay into multi-particle continua in the regime of strong interactions. 

Recent advances in the experimental technique of momentum-resolved electron energy-loss spectroscopy (M-EELS) \cite{MEELS} have made it possible to measure the dynamical charge response of numerous strongly correlated materials over a broad range of frequencies and momenta \cite{Abbamonte1,Abbamonte2,Abbamonte3}. 
Focusing specifically on the strange metal regime of a cuprate material (BSCCO), these experiments report evidence of a featureless particle-hole continuum extending over most of the Brillouin zone (BZ), while being independent of temperature and doping. Remarkably, the unconventional continuum persists up to the highest measurable energies and accounts for more than $99\%$ of the total spectral weight in the $f-$sum rule \cite{Abbamonte1,Abbamonte2}. Perhaps the most noteworthy observation is the absence of a sharply dispersing plasmon in the BZ (except for a narrow range of momenta, $q\lesssim0.05$ r.l.u., near the $\Gamma-$point \cite{Marel16,Hayden20}), as it decays into the featureless continuum. Evidence for such a continuum has been reported in earlier Raman studies \cite{bozovic87,ginsberg91} and  recent M-EELS measurements in other strongly interacting metals  (e.g. Sr$_2$RuO$_4$ \cite{Abbamonte3}). The microscopic origin for the decay of the plasmon into such continua remains unclear. Recent theoretical works have utilized solvable lattice electronic models \cite{DCrmp} to analyze the unconventional particle-hole continuum \cite{joshi} and the anomalous decay of plasmons \cite{XWDC} in the strongly correlated regime of certain non-Fermi liquid metals; see Ref.~\cite{zaanen} for a complementary holographic computation of plasmon decay.

\begin{figure*}[!ht]
\begin{center}
\includegraphics[width=1\linewidth]{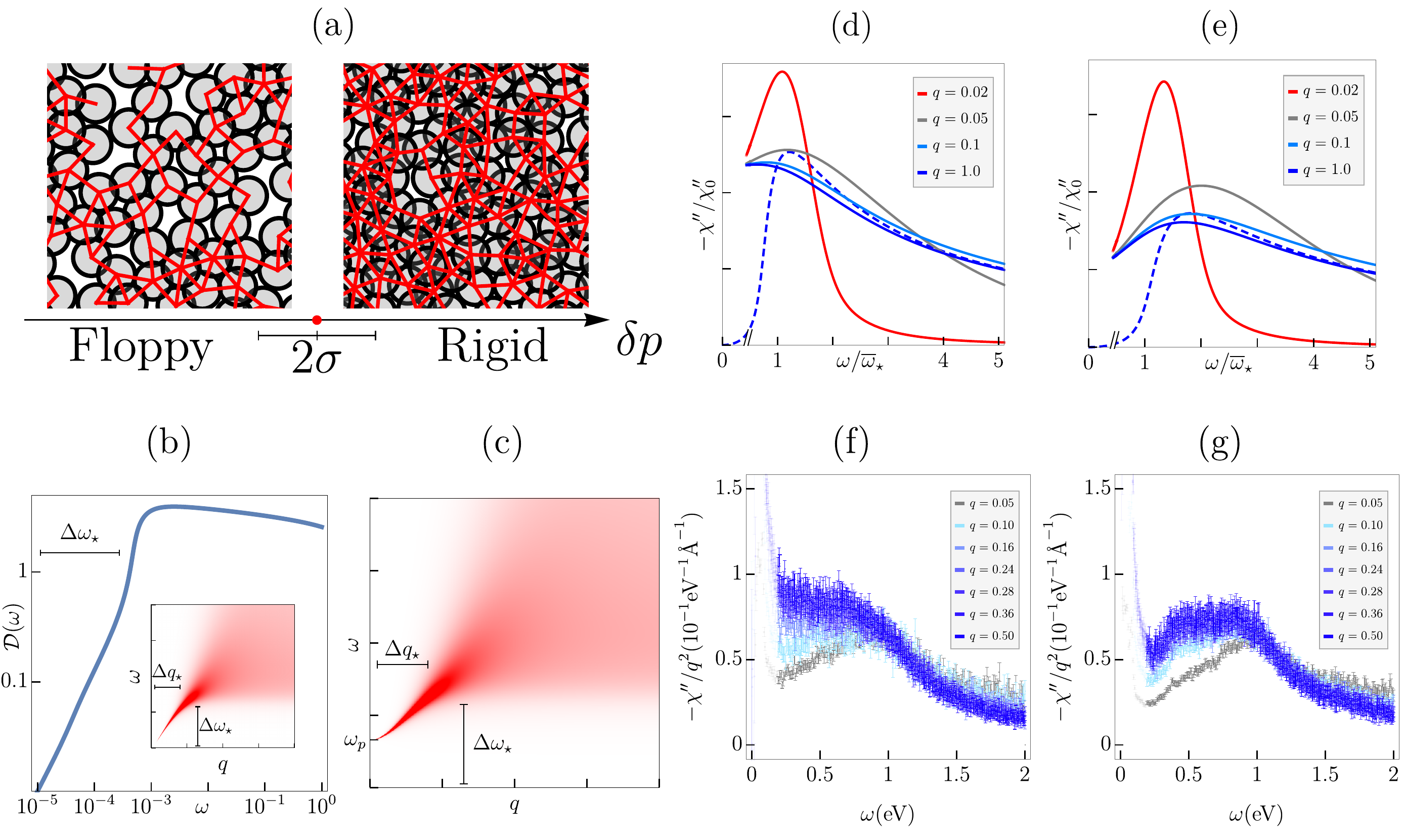}
\end{center}
\caption{{\bf Charge-density response near onset of rigidity  and in strange metals.} (a) A random network of bonds (red) displayed in a rigid vs. floppy system, on either side of a continuous rigidity percolation (RP) transition; the critical properties near RP are distinct from a jamming transition associated with random packings of hard spheres \cite{LiuNag2010}. We hypothesize that the two-particle density response over a broad range of intermediate energies near the hole-doping induced transition associated with the electrons near optimal doping in cuprates \cite{Keimer15} can be described as a rigidity-type transition. (b) The vibrational density of states, $\Dw$, as a function of frequency ($\omega$) at a fixed distance from the RP critical point $\DpRP=10^{-3}$. The plateau in $\Dw$ onsets for $\omega>\Delta\omega_\star\sim\left|\DpRP\right|$. Inset: The polarization function, $\Pi''(q,\omega)$, in the absence of Coulomb interactions, as a function of $q$ and $\omega$, revealing the acoustic collective mode and its damping inside the continuum. (c) The density response function, $\chi''(q,\omega)$, (including the Coulomb interaction) with  $\omega_p=0.66\left|\DpRP\right|$. The response functions, $\chi''(q,\omega)$, averaged over a range of $|\DpRP|\leq\sigma$ (see (a)) as a function of $\omega$ for different $q$ at a fixed distance ($\DpRP$) from RP for (d) $|\DpRP|=1.1\times10^{-3},~\sigma=0.9\times10^{-3}$, and (e) $|\DpRP|=1.6\times10^{-3},~ \sigma=0.9\times10^{-3}$. The plasma frequency is chosen to be at $\omega_p=0.5\,\overline{\omega}_\star$. The dashed line represents the $q$-independent shape of the imaginary part of the susceptibility in the absence of disorder. Frequencies in (d) and (e) are rescaled by the same scaling frequency $\overline{\omega}_\star$ associated with the average distance to the transition $\delta p$; see the Methods section for details. Experimental data from M-EELS \cite{Abbamonte1} demonstrating the overall $q$-independent shape of the continuum for (f) optimally doped BSCCO and (g) overdoped BSCCO. Error bars represent statistical (Poisson) error. The lowest frequencies show the lattice phonon, which we do not describe in our framework.}
\label{fig:ToyModel}
\end{figure*}

In addition to the anomalous dynamics of the charge density fluctuations, the normal metallic state across a number of strange metals exhibits universal scattering lifetimes \cite{Bruin13,APM21,zxs,johnson} and violates the Mott-Ioffe-Regel limit  with increasing temperature \cite{GunnarssonRMP,hussey}, suggesting an absence of electronic quasiparticles with a long mean-free path and lifetime. A satisfactory theoretical explanation for the complex and universal aspects of this phenomenology does not presently exist starting from microscopic models.

These results point to the intriguing possibility of the strongly interacting electron fluid forming a collective and self-organized, nearly jammed state. At intermediate energy scales, it is conceivable that certain aspects of the dynamical response associated with the collective modes can be understood by drawing analogies to a strongly correlated classical liquid. By analyzing the universal behavior of such a liquid near the onset of rigidity (Fig.~\ref{fig:ToyModel}(a)) and a detailed comparison to recent M-EELS experiments in cuprates, we conjecture that the intermediate-scale charge-dynamics in strange metals belongs in the family of a class of theories with critical rigidity correlations. This brings together a new class of problems under the umbrella of jamming, which includes rigidity transitions observed in colloids and granular materials \cite{LiuNag2010,BehringerCha2018}, living tissues \cite{manning16,heisenberg21}, elastic networks~\cite{FengGar1985,BroederszMac2011,LiarteLub2019,LiarteLub2020}, dislocation systems~\cite{MiguelZap2002,TsekenisGD11}, deep learning~\cite{BahriGan2020}, and analogues of metal-insulator transitions for interacting quantum bosons \cite{nussinov}. Quenched randomness in geometrically frustrated magnets has also been shown to produce a jammed spin liquid \cite{Bilitewski_2017}, which is known to display unconventional spin-dynamics \cite{Bilitewski_2019}.

In this manuscript we address the question of what phenomenon might give rise to a largely momentum-independent continuum such as that observed in M-EELS experiments. We conjecture that these observations might be connected to phenomena characteristic of the rigidity transition in granular media \cite{LiuNag2010}. Near such a transition, the vibrational density of states develops an anomalous, nearly frequency-independent plateau \cite{SilbertNag2005,Nagel09}. This paper will be concerned with addressing the similarities between the experimentally measured density correlations of strange metals and the calculated density correlations near the onset of rigidity, 
based on our recent analysis of the density response near a rigidity transition \cite{DL21a}.

\begin{center}
{\bf RESULTS}
\end{center}

The onset of rigidity in classical liquids (but without any long-range crystalline order) has a complex dynamical signature. The transition is associated with a singular rearrangement of the low-energy vibrational spectrum of the nearly rigid solid \cite{LiuNag2010}; see Fig.~\ref{fig:ToyModel}(b). These low-energy excitations will become the analog of the  unconventional particle-hole continuum in the strange metal that we described above. Moreover, the longitudinal phonons in these viscoelastic systems can decay into this continuum of low-energy vibrational excitations, much like the plasmons do in the cuprate strange metal.  

Starting with scaling forms for the longitudinal susceptibility that were derived recently by some of us \cite{DL21a,DL21b}, we will write down a coarse-grained effective description for the long-wavelength and low-frequency bosonic excitations in a liquid at the brink of rigidity percolation (RP); see Fig.~\ref{fig:ToyModel}(a). Percolation is a transition in connectivity; rigidity percolation is a transition from elastic to floppy, with a dramatic peak in low-energy excitations we believe common to strange metals.
We will extend our formalism to analyze the inelastic density-density response using the predictions of our scaling theory and make direct comparisons with the M-EELS results, highlighting the similarities between the mechanism for anomalous decay of the plasmon into the continuum at momenta away from the $\Gamma-$point. Given the relatively large energy-scales over which the charge response has been probed, it is likely that the quantum critical collective modes associated with various forms of broken symmetries that emerge at low-energies \cite{Keimer15} do not play a fundamental role in the interpretation of the M-EELS experiments. 

Our starting point is based on a recently proposed scaling ansatz for the dynamical susceptibilities near classical jamming and RP \cite{DL21a,DL21b}. There has been a dearth of solvable models in finite dimensions where universal features of the dynamical susceptibilities can be analyzed in a reliable fashion; we utilized the tractable effective medium theory \cite{LiarteLub2019} to compute these in Ref.~\cite{DL21a,DL21b} and obtained their explicit analytical forms. Given that the strange metals where the anomalous density fluctuations have been observed are quasi two-dimensional, we will model our system as a stack of weakly coupled two-dimensional layers. The individual layers are described in terms of a randomly percolated lattice of harmonic springs (Fig.~\ref{fig:ToyModel}(a)); the connection to the density fluctuations of an underlying electronic fluid will be made explicit later. For our present discussion, we will start specifically with the longitudinal part of the displacement response, $\Xi_L$, near RP,
\begin{subequations}
\beq
 &&\Xi_L(q,\omega)
        \approx -|\DpRP|^{-\gamma}\mathcal{L} \left(\widetilde{q},
            \widetilde{\omega}\right),
\label{eq:ResponseScaling}\\
&&\widetilde{q}\equiv  \frac{q}{|\DpRP|^{\nu}},~
\widetilde{\omega}\equiv\frac{\omega}{|\DpRP|^{z \nu}},
\eeq
\end{subequations}
where $q$ and $\omega$ represent the wavevector and frequency, respectively, and $|\delta p|$ represents the deviation away from the critical point. The critical exponents for susceptibility, correlation length, and correlation time are denoted $\gamma$, $\nu$ and $z$, respectively. For RP, our calculation leads to $\gamma=2$, $z=2$ and $\nu=1/2$. In two-dimensions, the above scaling form has additional dependence on the logarithms of the scaling variables which do not qualitatively affect any of our results; a detailed discussion of the origin of these additional logarithms will be discussed elsewhere (see Sec. I of the Supplementary Information \cite{SuppMaterial} for more details). $\mathcal{L}\left(\widetilde{q},
\widetilde{\omega}\right)$ is a universal scaling function whose explicit form appears in the Methods section. In all of our subsequent analysis and in our comparison with the experimental results, $\Xi_L(q,\omega)$ will play a central role. Near RP, the transverse response, $\Xi_T\left(q,\omega\right)$, has the same universal scaling form as $\Xi_L\left(q,\omega\right)$ but with different non-universal constants.

The onset of rigidity is tied to a significant rearrangement of the vibrational density of states, $\Dw$; see Fig.~\ref{fig:ToyModel}(b) and Methods for a definition. Near RP, $\Dw\sim\omega$ for $\omega\lesssim\D\omega_\star\sim|\DpRP|$ (up to additional logarithms).  
For $\omega\gtrsim\D\omega_\star$, $\Dw$ has a remarkably flat continuum as a function of $\omega$ over several orders of magnitude of frequencies; see Fig.~\ref{fig:ToyModel}(b). 
The physical origin of this low-energy continuum is related to the boson peak that demarcates a crossover from Debye to isostatic behavior, and is a recurring feature in the physics of glassy systems \cite{Nagel09,GiuliWya2014,FranzZam2015}. From the point of view of our analogy to the excitations in the strange metal, these modes are naturally interpreted as the particle-hole continuum. This analogy will become more direct when we analyze the nature of the collective excitations --- these are the phonons of the solid becoming floppy, which turn into the plasmon in the strange metal with the inclusion of Coulomb interactions --- and their decay into the flat $\Dw$ near RP.

In order to make the analogy between classical liquids and their vibrational excitations to the collective modes in strange metals, we need a precise relationship between the longitudinal susceptibility ($\Xi_{L}$) and the electron density correlation functions. As in the jellium model, we assume the negatively charged electronic liquid co-exists with a uniform oppositely charged (static) background to maintain electrical neutrality; we are only interested in the dynamics of the former. In the proposed model, the changes in the local displacement, ${\cal{U}}$, are tied to a local fluctuation of the electronic number density. More precisely,
\beq
n(\x)=n_0(1-\nabla\cdot{\cal{U}}),
\eeq
where $n_0=\rho/m$ is the average background density. One of the central quantities of interest is the polarization function, $\Pi(q,\omega) = n_0^2q^2~\Xi_L(q,\omega)$, which is related to the longitudinal susceptibility introduced earlier. This is the density-density response of the neutral system near the transition. Since the electronic liquid is charged and interacts via repulsive Coulomb interactions, $V(|\x-\x'|)$, we include it explicitly as
\beq
\Delta U = \frac{1}{2}\int_{\x}\int_{\x'} \delta n(x) V(|\x-\x'|) \delta n(x'),
\eeq
where $\delta n(\x) = n(\x)-n_0 = -n_0\nabla\cdot{\cal{U}}$. The experimentally measured density-density response, $\chi(q,\omega)$, can be obtained from the polarizability after including the effects of Coulomb interactions,
\beq
\chi(q,\omega) = \frac{\Pi(q,\omega)}{1-V(q)\Pi(q,\omega)}. 
\eeq
In the remainder of this study, we will calculate $\chi(q,\omega)$ near RP using the universal form of $\Xi_L(q,\omega)$, and highlighting both its similarities and differences when compared against the experimentally measured density response function in the cuprate strange metal. See Sec. II of the Supplementary Information \cite{SuppMaterial} for more details.


To analyze the effect of the plasmon decay into the continuum, it is conceptually simpler to approach the transition from the rigid side. The imaginary part of the susceptibility, $\chi''(q,\omega)$, reveals a sharply dispersing plasmon for $\D q_\star\sim\left|\DpRP\right|^{1/2}$ (up to logarithms), controlled by the distance to RP ($\DpRP$), that broadens significantly as a result of decay into the low-energy vibrational states over a broad range of wavevectors and frequencies; see inset of Fig.~\ref{fig:ToyModel}(b). The effect of $V(q)$ on $\chi(q,\omega)$ is to renormalize the acoustic mode to the plasma frequency, $\omega_p=\sqrt{4\pi e^2 n_0/m}$, where we have assumed the three-dimensional form, $V(q)=4\pi e^2/q^2$; see Fig.~\ref{fig:ToyModel}(c). The broadening of the plasmon due to decay into the unconventional continuum remains identical. The phenomenology described here is exactly what we set out to achieve theoretically inspired by the M-EELS experiments in strange metals---a plasmon that is damped beyond small momenta $q\gtrsim\Delta q_\star$ into a featureless, low-energy continuum. The close similarity that we demonstrate between the unconventional decay of the phonon into the vibrational continuum near RP and of the plasmon into the measured particle-hole continuum in strange metals is one of the central results of this paper.

\begin{figure}[!ht]
\begin{center}
\includegraphics[width=\linewidth]{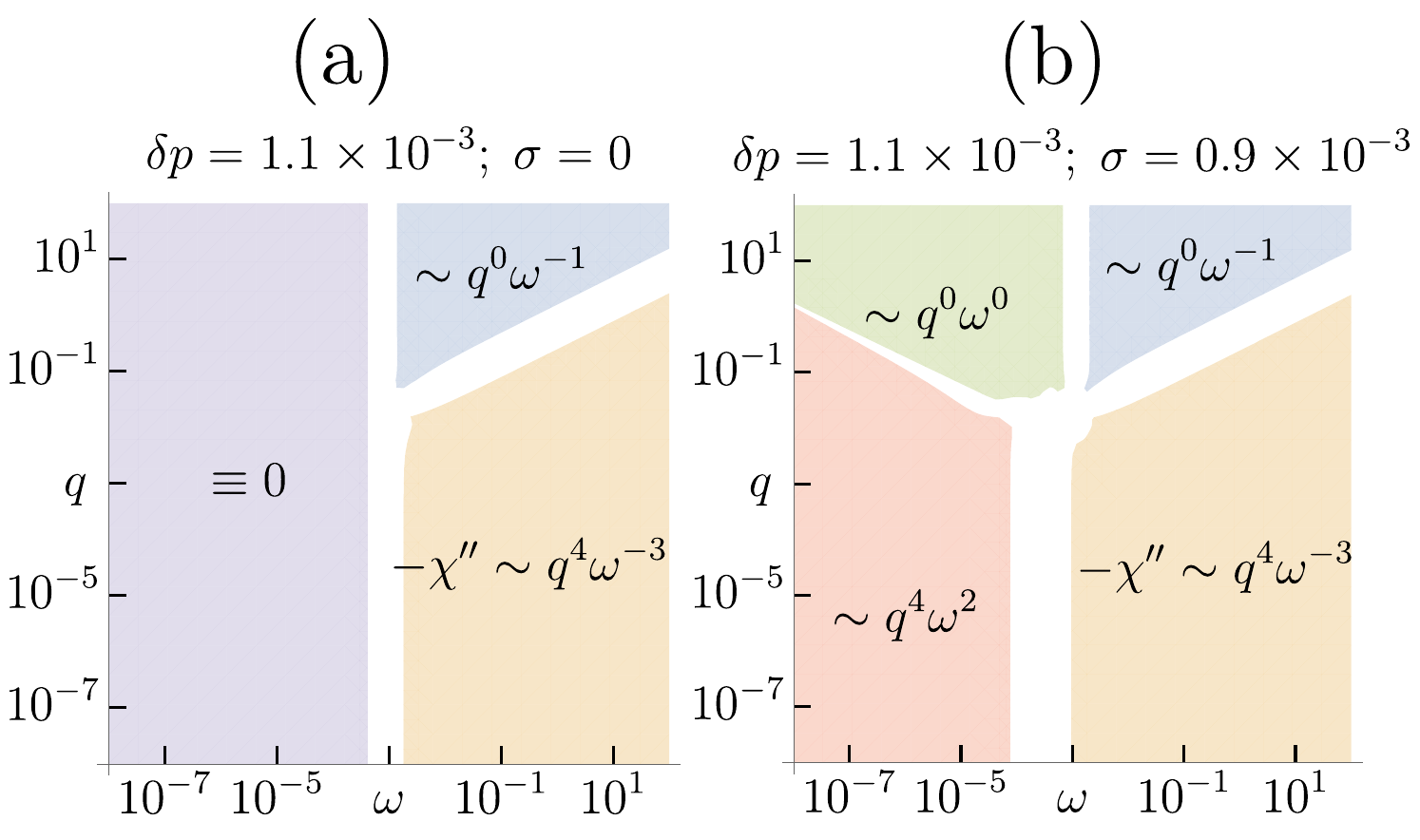}
\end{center}
\caption{\textbf{Distinct spectroscopic regimes of the charge-density response.} Frequency and momentum dependence of $\chi''$, (a) without ($\sigma=0$) and (b) with ($\sigma=0.9\times10^{-3}$) disorder averaging. Here $\left|\DpRP\right|=1.1\times10^{-3}$, $e=10^{-4}$. In (a), the largest values of $q$ lead to a bump in the susceptibility at a $q$-independent frequency $\Delta\omega_\star$, followed by a decay $\sim\omega^{-1}$ eventually crossing over into $\sim\omega^{-3}$ Drude-like behavior. In (b), a plateau in the response emerges at the lowest frequencies whose width is set by $\sigma$. The qualitative behavior is retained even after including corrections to the response that fix the scaling in the lowest frequency regime (see Fig.~\ref{fig:ToyModel}(d),(e)).}
\label{fig:PowerLaws}
\end{figure}

Let us next turn to studying the detailed $q,\omega-$dependence of $\chi(q,\omega)$ near RP in order to make further comparisons with the measured charge response functions. For the smallest values of $q$, there is a sharp plasmon that appears at the plasma frequency, $\omega_p$. For $q\gtrsim \Delta q_{\star}\sim\left|\DpRP\right|^{1/2}$, the plasmon broadens rapidly, and $\chi(q,\omega)$ becomes nearly $q$-independent with a broad feature centered near $\Delta\omega_{\star}$. Increasing $q$ further serves only to adjust the crossover frequency beyond which there is a crossover to a $1/\omega^3$ falloff, in accordance with the \textit{f}-sum rule (see Figure~\ref{fig:PowerLaws}(a)). The $q$-independent shape of $\chi''(q,\omega)$ is also shown as the dashed blue curve in Figure~\ref{fig:ToyModel}(d)-(e). This broad feature is tied to the same boson peak that was discussed above in the context of the onset of the enhancement of the low-energy modes in $\mathcal{D}(\omega)$.

Although our form of $\chi''\left(q,\omega\right)$ near the transition reproduces the strongly overdamped plasmon and the $q$-independent shape of the response over the measured frequency range, the response at the lowest frequencies does not have the characteristic plateau of the experiment. To address the possible origin of this feature, we can appeal to the inherent inhomogeneity that is present in these materials. There is experimental evidence for nanoscale electronic inhomogeneity across multiple families of cuprate single crystals (including, e.g. BSCCO) \cite{disorder1,disorder2}. For a given sample at a fixed nominal doping, the experiments probe the density response averaged over all of the inhomogeneous regions of the sample. To replicate this feature in our theoretical analysis, we sample and smear our results for $\chi(q,\omega)$ over a distribution of $\DpRP$. We thus assume that the variations in doping level change the distance to the onset of rigidity. Our averaging presumes the disorder does not couple to the translational Goldstone mode of the transition (by solely changing the density of bonds). The doping, however, breaks translational symmetry, and pinning on defects is also known to lower the threshold of rigidity ~\cite{ReichhardtGNR12,GravesNPGLS16,peter2018crossover,ZhangRPSBVUBG22}. Adding the effects of pinning to our analysis could be fruitful in future work.
The qualitative effects of the above averaging procedure are similar for any smooth, symmetric distribution. 

Near the boson peak, the disorder-averaged susceptibility is most drastically altered. When the mean deviation from criticality is comparable to the width of the disorder distribution, $|\DpRP|\leq\sigma$, the spectrum becomes dispersionless as a function of $\omega$ for large $q>\Delta q_\star$; see Fig.~\ref{fig:ToyModel}(d). Within our framework, the frequency-independent plateau observed near optimal doping can be interpreted as the disorder-induced smearing of the boson peak near RP. Beyond this featureless region, there is a crossover into an anomalous power-law regime, $\chi''(q,\omega)\sim 1/\omega^{\alpha}$ with $\alpha<3$. Both of these features are similar to the experiments \cite{Abbamonte1}; see Fig.~\ref{fig:ToyModel}(d)-(g) for a comparison. At the largest frequencies, the asymptotic forms of the polarization with and without disorder-averaging are identical with $\alpha=3$. The shape of the susceptibility is largely independent of $q$ over a wide range of $\omega$. This leads to a $q$-independent crossover frequency from the plateau to a power-law fall off at large $\omega$. As we move away from the transition fixing the magnitude of the disorder $\sigma$, the plateau at low frequencies evolves into a bump; see Fig.~\ref{fig:ToyModel}(e). This bump can be interpreted as a severely overdamped plasmon, whose location  becomes nearly $q$-independent at large $q$. The $q$-independence is tied to the decay into the particle-hole continuum, whose onset is at a fixed $\Delta\omega_\star$.

The power-law scaling behavior of the singular part of the susceptibility $\chi''$ before and after the inclusion of disorder averaging is illustrated in Fig.~\ref{fig:PowerLaws}. At the highest frequencies, the power-law scaling is unaffected by the specific type of disorder considered here. At low frequencies, we see the emergence of a plateau region whose width is $q$-independent (and set by the amount of disorder $\sigma$) for $q\gtrsim\Delta q_\star$. For experimental measurements close to this critical point, all wavevectors except for those closest to the center of the BZ will probe the incoherent plateau rather than the collective mode. The most notable difference between this framework and the one observed in the experiments is in the wavevector dependence of the magnitude of the response. If the response has a $q$-independent shape at all frequencies, then one infers that it must scale as $\sim q^2$ to satisfy the \textit{f}-sum rule. The singular responses computed in this paper also satisfy the appropriate sum rules, since $q$ sets the frequency at which we cross over into the Drude-type scaling $\sim\omega^{-3}$. See Sec. III of the Supplementary Information \cite{SuppMaterial} for more details. A recent complimentary theoretical work \cite{XWDC} finds a distinct high-frequency scaling $\sim1/\omega^2\log^2(\omega)$, which is also consistent with the $f$-sum rule and is in better qualitative agreement with the experiments.

Within the framework of rigidity percolation, we have pointed out an intriguing analogy between the  large collection of low-energy vibrational modes and the particle-hole continuum of strange metals, into which collective modes can rapidly decay. The onset energy of this decay is set by the distance to the critical point. Although the details of the specific momentum-dependence for the polarizability are not in  perfect agreement with the MEELS experiment, we can reproduce a $q$-independent shape for $q>\Delta q_{\star}$ that is set by the distance to the critical point. It is possible that a different,  and yet to be understood, universality class of rigidity transition displays a power-law density response that agrees  better with the experiments. 
A broad implication of our hypothesis is that over a range of intermediate energy scales over which the density correlations in strange metals appear to display features like jamming, the electronic fluid might also display interesting memory effects known to arise in glassy systems and near rigidity transitions. Finding new experimental ways to probe this physics remains an interesting future direction.
Developing a microscopic quantum theory of interacting electrons  whose effective theory reduces to an analogous rigidity transition is a challenging open problem \cite{schmalian,nussinov}. In this regard, exploring possible connections between the low-energy vibrational excitations near jamming and the low-energy non-quasiparticle-like excitations in the solvable quantum Sachdev-Ye-Kitaev models \cite{DCrmp,kurchan} will be an interesting theoretical exercise.

\begin{center}
{\bf METHODS}
\end{center}
\begin{center}
{\bf Universal scaling function}
\end{center}

In three spatial dimensions and higher, the scaling function $\mathcal{L}(\widetilde{q},\widetilde{\omega})$ has the explicit form \cite{DL21a,DL21b}
\begin{subequations}
\beq
\mathcal{L}(\widetilde{q},\widetilde{\omega})
        &=& \left[a \, \widetilde{q}^2 \mathcal{M}_{\pm} (\widetilde{\omega})
            - \widetilde{v} (\widetilde{\omega})\right]^{-1},\label{eq:LbarDef}\\
\mathcal{M}_{\pm} (\widetilde{\omega}) &=& b \left[\sqrt{1 - c \, \widetilde{v}(\widetilde{\omega})} \pm 1\right],
\eeq
\end{subequations}
where $a,~b$ and $c$ are constants, with $\widetilde{v} (\widetilde{\omega}) = \rho \,\widetilde{\omega}^2$
and $i \, \gamma \, \widetilde{\omega}$ for undamped and overdamped dynamics, respectively. Here $\rho$ is a mass density and $\gamma$ is a viscous drag coefficient. The plus and minus signs in $\mathcal{M}_{\pm}(...)$ correspond to 
the rigid and floppy states, respectively. We use the undamped form of the response exclusively.
\begin{center}
{\bf Vibrational density of states}
\end{center}
The vibrational density of states can be computed from
\begin{subequations}
\beq
\Dw &=& -\frac{\omega}{\pi}\int_{\tn{BZ}}d^2q\; \textrm{Tr}\left(\textrm{Im}\left[\mathcal{G}_{ij}\left(q,\omega\right)\right]\right), \\
\mathcal{G}_{ij}\left(q,\omega\right)&=&\Xi_L\left(q,\omega\right)\hat{q}_i\hat{q}_j+\Xi_T\left(q,\omega\right)\left(\delta_{ij}-\hat{q}_i\hat{q}_j\right)
\eeq
\end{subequations}
The density of states has an additional $\omega$ prefactor as we are considering excitations in a classical disordered system. 

\begin{center}
{\bf Disorder averaging of charge-density response}
\end{center}

 We convolve our universal scaling function with a specific disorder distribution, such that the effective disorder-averaged polarization function takes the form (denoted `$\overline{~~}$')
\begin{subequations}
\beq
\overline{\chi_{\DpRP}(q,\omega)} &=& 
\int_{-\infty}^{\infty}d\left(\D\pRP'\right)  {\cal{P}}_\sigma(\D\pRP')~ \chi_{\DpRP'}(q,\omega)\\
{\cal{P}}_\sigma[\D\pRP'] &=&\frac{1}{\sqrt{2\pi\sigma^2}}e^{-\left(\D\pRP'\right)^2/2\sigma^2},
\eeq
\end{subequations}
where $\chi_{\DpRP}(q,\omega)$ is the response at a fixed distance ($\DpRP$) from RP, and we choose a Gaussian distribution, ${\cal{P}}_\sigma[\D\pRP]$, with width $\sigma$ and $\D\pRP'\equiv\DpRP'-\DpRP$. For the forms of the response in Figures~\ref{fig:ToyModel}(d)-(e), the frequency is rescaled by $\overline{\omega}_{\star}=10^{-3}$ in both figures, instead of the distinct $\DpRP$. This is to make comparison to the experiment, where the dopant concentration is changed (moving the system further from a critical point in our framework), but not the frequency scale. 
\bigbreak

\begin{center}
{\bf ACKNOWLEDGEMENTS}
\end{center}

P.A. thanks Vincenzo Vitelli for  suggesting a possible connection between jamming and the measured density response in Fig.~\ref{fig:ToyModel}. This  work  was  supported  in part by NSF DMR-1719490 (SJT and JPS). DC is supported by faculty startup funds at Cornell University. P.A. acknowledges support from the EPiQS program of the Gordon and Betty Moore Foundation, grant no. GBMF9452.
DBL thanks financial support through FAPESP grants \verb|#| 2016/01343-7 and \verb|#| 2021/14285-3.

\bibliographystyle{unsrtnat}
\normalem
\bibliography{refscombV4}

\begin{thebibliography}{52}
\providecommand{\natexlab}[1]{#1}
\providecommand{\url}[1]{\texttt{#1}}
\expandafter\ifx\csname urlstyle\endcsname\relax
  \providecommand{\doi}[1]{doi: #1}\else
  \providecommand{\doi}{doi: \begingroup \urlstyle{rm}\Url}\fi

\bibitem[Pines and Nozi\`eres(2018)]{pines}
David Pines and Philippe Nozi\`eres.
\newblock \emph{Theory of Quantum Liquids: Normal Fermi Liquids}.
\newblock CRC Press, 2018.

\bibitem[Vig et~al.(2017)Vig, Kogar, Mitrano, Husain, Mishra, Rak, Venema,
  Johnson, Gu, Fradkin, Norman, and Abbamonte]{MEELS}
Sean Vig, Anshul Kogar, Matteo Mitrano, Ali~A. Husain, Vivek Mishra, Melinda~S.
  Rak, Luc Venema, Peter~D. Johnson, Genda~D. Gu, Eduardo Fradkin, Michael~R.
  Norman, and Peter Abbamonte.
\newblock {Measurement of the dynamic charge response of materials using
  low-energy, momentum-resolved electron energy-loss spectroscopy (M-EELS)}.
\newblock \emph{SciPost Phys.}, 3:\penalty0 026, 2017.
\newblock \doi{10.21468/SciPostPhys.3.4.026}.
\newblock URL \url{https://scipost.org/10.21468/SciPostPhys.3.4.026}.

\bibitem[Mitrano et~al.(2018)Mitrano, Husain, Vig, Kogar, Rak, Rubeck,
  Schmalian, Uchoa, Schneeloch, Zhong, Gu, and Abbamonte]{Abbamonte1}
M.~Mitrano, A.~A. Husain, S.~Vig, A.~Kogar, M.~S. Rak, S.~I. Rubeck,
  J.~Schmalian, B.~Uchoa, J.~Schneeloch, R.~Zhong, G.~D. Gu, and P.~Abbamonte.
\newblock Anomalous density fluctuations in a strange metal.
\newblock \emph{Proceedings of the National Academy of Sciences}, 115\penalty0
  (21):\penalty0 5392--5396, 2018.
\newblock ISSN 0027-8424.
\newblock \doi{10.1073/pnas.1721495115}.
\newblock URL \url{https://www.pnas.org/content/115/21/5392}.

\bibitem[Husain et~al.(2019)Husain, Mitrano, Rak, Rubeck, Uchoa, March, Dwyer,
  Schneeloch, Zhong, Gu, and Abbamonte]{Abbamonte2}
Ali~A. Husain, Matteo Mitrano, Melinda~S. Rak, Samantha Rubeck, Bruno Uchoa,
  Katia March, Christian Dwyer, John Schneeloch, Ruidan Zhong, G.~D. Gu, and
  Peter Abbamonte.
\newblock Crossover of charge fluctuations across the strange metal phase
  diagram.
\newblock \emph{Phys. Rev. X}, 9:\penalty0 041062, Dec 2019.
\newblock \doi{10.1103/PhysRevX.9.041062}.
\newblock URL \url{https://link.aps.org/doi/10.1103/PhysRevX.9.041062}.

\bibitem[{Husain} et~al.(2020){Husain}, {Huang}, {Mitrano}, {Rak}, {Rubeck},
  {Guo}, {Yang}, {Sow}, {Maeno}, {Uchoa}, {Chiang}, {Batson}, {Phillips}, and
  {Abbamonte}]{Abbamonte3}
A.~A. {Husain}, E.~W. {Huang}, M.~{Mitrano}, M.~S. {Rak}, S.~I. {Rubeck},
  X.~{Guo}, H.~{Yang}, C.~{Sow}, Y.~{Maeno}, B.~{Uchoa}, T.~C. {Chiang}, P.~E.
  {Batson}, P.~W. {Phillips}, and P.~{Abbamonte}.
\newblock {Observation of Pines' Demon in Sr$_2$RuO$_4$}.
\newblock \emph{arXiv e-prints}, art. arXiv:2007.06670, July 2020.

\bibitem[Levallois et~al.(2016)Levallois, Tran, Pouliot, Presura, Greene,
  Eckstein, Uccelli, Giannini, Gu, Leggett, and van~der Marel]{Marel16}
J.~Levallois, M.~K. Tran, D.~Pouliot, C.~N. Presura, L.~H. Greene, J.~N.
  Eckstein, J.~Uccelli, E.~Giannini, G.~D. Gu, A.~J. Leggett, and D.~van~der
  Marel.
\newblock Temperature-dependent ellipsometry measurements of partial coulomb
  energy in superconducting cuprates.
\newblock \emph{Phys. Rev. X}, 6:\penalty0 031027, Aug 2016.
\newblock \doi{10.1103/PhysRevX.6.031027}.
\newblock URL \url{https://link.aps.org/doi/10.1103/PhysRevX.6.031027}.

\bibitem[Nag et~al.(2020)Nag, Zhu, Bejas, Li, Robarts, Yamase, Petsch, Song,
  Eisaki, Walters, Garc\'{\i}a-Fern\'andez, Greco, Hayden, and Zhou]{Hayden20}
Abhishek Nag, M.~Zhu, Mat\'{\i}as Bejas, J.~Li, H.~C. Robarts, Hiroyuki Yamase,
  A.~N. Petsch, D.~Song, H.~Eisaki, A.~C. Walters, M.~Garc\'{\i}a-Fern\'andez,
  Andr\'es Greco, S.~M. Hayden, and Ke-Jin Zhou.
\newblock Detection of acoustic plasmons in hole-doped lanthanum and bismuth
  cuprate superconductors using resonant inelastic x-ray scattering.
\newblock \emph{Phys. Rev. Lett.}, 125:\penalty0 257002, Dec 2020.
\newblock \doi{10.1103/PhysRevLett.125.257002}.
\newblock URL \url{https://link.aps.org/doi/10.1103/PhysRevLett.125.257002}.

\bibitem[Bozovic et~al.(1987)Bozovic, Kirillov, Kapitulnik, Char, Hahn,
  Beasley, Geballe, Kim, and Heeger]{bozovic87}
I.~Bozovic, D.~Kirillov, A.~Kapitulnik, K.~Char, M.~R. Hahn, M.~R. Beasley,
  T.~H. Geballe, Y.~H. Kim, and A.~J. Heeger.
\newblock Optical measurements on oriented thin
  ${\mathrm{yba}}_{2}$${\mathrm{cu}}_{3}$${\mathrm{o}}_{7\mathrm{\ensuremath{-}}\mathrm{\ensuremath{\delta}}}$
  films: Lack of evidence for excitonic superconductivity.
\newblock \emph{Phys. Rev. Lett.}, 59:\penalty0 2219--2221, Nov 1987.
\newblock \doi{10.1103/PhysRevLett.59.2219}.
\newblock URL \url{https://link.aps.org/doi/10.1103/PhysRevLett.59.2219}.

\bibitem[Slakey et~al.(1991)Slakey, Klein, Rice, and Ginsberg]{ginsberg91}
F.~Slakey, M.~V. Klein, J.~P. Rice, and D.~M. Ginsberg.
\newblock Raman investigation of the
  ${\mathrm{yba}}_{2}$${\mathrm{cu}}_{3}$${\mathrm{o}}_{7}$ imaginary response
  function.
\newblock \emph{Phys. Rev. B}, 43:\penalty0 3764--3767, Feb 1991.
\newblock \doi{10.1103/PhysRevB.43.3764}.
\newblock URL \url{https://link.aps.org/doi/10.1103/PhysRevB.43.3764}.

\bibitem[Chowdhury et~al.(2022)Chowdhury, Georges, Parcollet, and
  Sachdev]{DCrmp}
Debanjan Chowdhury, Antoine Georges, Olivier Parcollet, and Subir Sachdev.
\newblock Sachdev-ye-kitaev models and beyond: Window into non-fermi liquids.
\newblock \emph{Rev. Mod. Phys.}, 94:\penalty0 035004, Sep 2022.
\newblock \doi{10.1103/RevModPhys.94.035004}.
\newblock URL \url{https://link.aps.org/doi/10.1103/RevModPhys.94.035004}.

\bibitem[Joshi and Sachdev(2020)]{joshi}
Darshan~G. Joshi and Subir Sachdev.
\newblock Anomalous density fluctuations in a random $t\text{\ensuremath{-}}j$
  model.
\newblock \emph{Phys. Rev. B}, 102:\penalty0 165146, Oct 2020.
\newblock \doi{10.1103/PhysRevB.102.165146}.
\newblock URL \url{https://link.aps.org/doi/10.1103/PhysRevB.102.165146}.

\bibitem[Wang and Chowdhury(2023)]{XWDC}
Xuepeng Wang and Debanjan Chowdhury.
\newblock Collective density fluctuations of strange metals with critical fermi
  surfaces.
\newblock \emph{Phys. Rev. B}, 107:\penalty0 125157, Mar 2023.
\newblock \doi{10.1103/PhysRevB.107.125157}.
\newblock URL \url{https://link.aps.org/doi/10.1103/PhysRevB.107.125157}.

\bibitem[Romero-Berm\'udez et~al.(2019)Romero-Berm\'udez, Krikun, Schalm, and
  Zaanen]{zaanen}
Aurelio Romero-Berm\'udez, Alexander Krikun, Koenraad Schalm, and Jan Zaanen.
\newblock Anomalous attenuation of plasmons in strange metals and holography.
\newblock \emph{Phys. Rev. B}, 99:\penalty0 235149, Jun 2019.
\newblock \doi{10.1103/PhysRevB.99.235149}.
\newblock URL \url{https://link.aps.org/doi/10.1103/PhysRevB.99.235149}.

\bibitem[Liu and Nagel(2010)]{LiuNag2010}
Andrea~J. Liu and Sidney~R. Nagel.
\newblock The jamming transition and the marginally jammed solid.
\newblock \emph{Annual Review of Condensed Matter Physics}, 1\penalty0
  (1):\penalty0 347--369, 2010.
\newblock \doi{10.1146/annurev-conmatphys-070909-104045}.
\newblock URL \url{https://doi.org/10.1146/annurev-conmatphys-070909-104045}.

\bibitem[Keimer et~al.(2015)Keimer, Kivelson, Norman, Uchida, and
  Zaanen]{Keimer15}
B.~Keimer, S.~A. Kivelson, M.~R. Norman, S.~Uchida, and J.~Zaanen.
\newblock From quantum matter to high-temperature superconductivity in copper
  oxides.
\newblock \emph{Nature}, 518\penalty0 (7538):\penalty0 179--186, Feb 2015.
\newblock ISSN 0028-0836.
\newblock URL \url{http://dx.doi.org/10.1038/nature14165}.

\bibitem[Bruin et~al.(2013)Bruin, Sakai, Perry, and Mackenzie]{Bruin13}
J.~A.~N. Bruin, H.~Sakai, R.~S. Perry, and A.~P. Mackenzie.
\newblock Similarity of scattering rates in metals showing $t-$linear
  resistivity.
\newblock \emph{Science}, 339\penalty0 (6121):\penalty0 804--807, 2013.
\newblock ISSN 0036-8075.

\bibitem[{Hartnoll} and {Mackenzie}(2021)]{APM21}
Sean~A. {Hartnoll} and Andrew~P. {Mackenzie}.
\newblock {Planckian Dissipation in Metals}.
\newblock \emph{arXiv e-prints}, art. arXiv:2107.07802, July 2021.

\bibitem[Damascelli et~al.(2003)Damascelli, Hussain, and Shen]{zxs}
Andrea Damascelli, Zahid Hussain, and Zhi-Xun Shen.
\newblock Angle-resolved photoemission studies of the cuprate superconductors.
\newblock \emph{Rev. Mod. Phys.}, 75:\penalty0 473--541, Apr 2003.
\newblock \doi{10.1103/RevModPhys.75.473}.
\newblock URL \url{https://link.aps.org/doi/10.1103/RevModPhys.75.473}.

\bibitem[Wang et~al.(2004)Wang, Yang, Sekharan, Ding, Engelbrecht, Dai, Wang,
  Kaminski, Valla, Kidd, Fedorov, and Johnson]{johnson}
S.-C. Wang, H.-B. Yang, A.~K.~P. Sekharan, H.~Ding, J.~R. Engelbrecht, X.~Dai,
  Z.~Wang, A.~Kaminski, T.~Valla, T.~Kidd, A.~V. Fedorov, and P.~D. Johnson.
\newblock Quasiparticle line shape of ${\mathrm{sr}}_{2}{\mathrm{ruo}}_{4}$ and
  its relation to anisotropic transport.
\newblock \emph{Phys. Rev. Lett.}, 92:\penalty0 137002, Apr 2004.
\newblock \doi{10.1103/PhysRevLett.92.137002}.
\newblock URL \url{https://link.aps.org/doi/10.1103/PhysRevLett.92.137002}.

\bibitem[Gunnarsson et~al.(2003)Gunnarsson, Calandra, and Han]{GunnarssonRMP}
O.~Gunnarsson, M.~Calandra, and J.~E. Han.
\newblock Colloquium: Saturation of electrical resistivity.
\newblock \emph{Rev. Mod. Phys.}, 75:\penalty0 1085--1099, Oct 2003.
\newblock \doi{10.1103/RevModPhys.75.1085}.
\newblock URL \url{https://link.aps.org/doi/10.1103/RevModPhys.75.1085}.

\bibitem[Hussey et~al.(2004)Hussey, Takenaka, and Takagi]{hussey}
N.~E. Hussey, K.~Takenaka, and H.~Takagi.
\newblock Universality of the mott–ioffe–regel limit in metals.
\newblock \emph{Philosophical Magazine}, 84\penalty0 (27):\penalty0 2847--2864,
  2004.
\newblock \doi{10.1080/14786430410001716944}.
\newblock URL \url{https://doi.org/10.1080/14786430410001716944}.

\bibitem[Behringer and Chakraborty(2018)]{BehringerCha2018}
Robert~P Behringer and Bulbul Chakraborty.
\newblock The physics of jamming for granular materials: a review.
\newblock \emph{Reports on Progress in Physics}, 82\penalty0 (1):\penalty0
  012601, nov 2018.
\newblock \doi{10.1088/1361-6633/aadc3c}.
\newblock URL \url{https://doi.org/10.1088/1361-6633/aadc3c}.

\bibitem[Bi et~al.(2016)Bi, Yang, Marchetti, and Manning]{manning16}
Dapeng Bi, Xingbo Yang, M.~Cristina Marchetti, and M.~Lisa Manning.
\newblock Motility-driven glass and jamming transitions in biological tissues.
\newblock \emph{Phys. Rev. X}, 6:\penalty0 021011, Apr 2016.
\newblock \doi{10.1103/PhysRevX.6.021011}.
\newblock URL \url{https://link.aps.org/doi/10.1103/PhysRevX.6.021011}.

\bibitem[Petridou et~al.(2021)Petridou, Corominas-Murtra, Heisenberg, and
  Hannezo]{heisenberg21}
Nicoletta~I. Petridou, Bernat Corominas-Murtra, Carl-Philipp Heisenberg, and
  Edouard Hannezo.
\newblock Rigidity percolation uncovers a structural basis for embryonic tissue
  phase transitions.
\newblock \emph{Cell}, 184\penalty0 (7):\penalty0 1914--1928.e19, 2021.
\newblock ISSN 0092-8674.
\newblock \doi{https://doi.org/10.1016/j.cell.2021.02.017}.
\newblock URL
  \url{https://www.sciencedirect.com/science/article/pii/S0092867421001677}.

\bibitem[Feng et~al.(1985)Feng, Thorpe, and Garboczi]{FengGar1985}
S.~Feng, M.~F. Thorpe, and E.~Garboczi.
\newblock Effective-medium theory of percolation on central-force elastic
  networks.
\newblock \emph{Physical Review B}, 31\penalty0 (1):\penalty0 276--280, 1985.
\newblock URL \url{<Go to ISI>://A1985TZ38500031}.

\bibitem[Broedersz et~al.(2011)Broedersz, Mao, Lubensky, and
  MacKintosh]{BroederszMac2011}
Chase~P Broedersz, Xiaoming Mao, Tom~C Lubensky, and Frederick~C MacKintosh.
\newblock Criticality and isostaticity in fibre networks.
\newblock \emph{Nature Physics}, 7\penalty0 (12):\penalty0 983--988, 2011.

\bibitem[Liarte et~al.(2019)Liarte, Mao, Stenull, and Lubensky]{LiarteLub2019}
Danilo~B. Liarte, Xiaoming Mao, Olaf Stenull, and T.~C. Lubensky.
\newblock Jamming as a multicritical point.
\newblock \emph{Phys. Rev. Lett.}, 122:\penalty0 128006, Mar 2019.
\newblock \doi{10.1103/PhysRevLett.122.128006}.
\newblock URL \url{https://link.aps.org/doi/10.1103/PhysRevLett.122.128006}.

\bibitem[Liarte et~al.(2020)Liarte, Stenull, and Lubensky]{LiarteLub2020}
Danilo~B. Liarte, O.~Stenull, and T.~C. Lubensky.
\newblock Multifunctional twisted kagome lattices: Tuning by pruning mechanical
  metamaterials.
\newblock \emph{Phys. Rev. E}, 101:\penalty0 063001, Jun 2020.
\newblock \doi{10.1103/PhysRevE.101.063001}.
\newblock URL \url{https://link.aps.org/doi/10.1103/PhysRevE.101.063001}.

\bibitem[Miguel et~al.(2002)Miguel, Vespignani, Zaiser, and
  Zapperi]{MiguelZap2002}
M.-Carmen Miguel, Alessandro Vespignani, Michael Zaiser, and Stefano Zapperi.
\newblock Dislocation jamming and andrade creep.
\newblock \emph{Phys. Rev. Lett.}, 89:\penalty0 165501, Sep 2002.
\newblock \doi{10.1103/PhysRevLett.89.165501}.
\newblock URL \url{https://link.aps.org/doi/10.1103/PhysRevLett.89.165501}.

\bibitem[Tsekenis et~al.(2011)Tsekenis, Goldenfeld, and Dahmen]{TsekenisGD11}
Georgios Tsekenis, Nigel Goldenfeld, and Karin~A Dahmen.
\newblock Dislocations jam at any density.
\newblock \emph{Physical review letters}, 106\penalty0 (10):\penalty0 105501,
  2011.

\bibitem[Bahri et~al.(2020)Bahri, Kadmon, Pennington, Schoenholz,
  Sohl-Dickstein, and Ganguli]{BahriGan2020}
Yasaman Bahri, Jonathan Kadmon, Jeffrey Pennington, Sam~S. Schoenholz, Jascha
  Sohl-Dickstein, and Surya Ganguli.
\newblock Statistical mechanics of deep learning.
\newblock \emph{Annual Review of Condensed Matter Physics}, 11\penalty0
  (1):\penalty0 501--528, 2020.
\newblock \doi{10.1146/annurev-conmatphys-031119-050745}.
\newblock URL \url{https://doi.org/10.1146/annurev-conmatphys-031119-050745}.

\bibitem[Nussinov et~al.(2013)Nussinov, Johnson, Graf, and Balatsky]{nussinov}
Zohar Nussinov, Patrick Johnson, Matthias~J. Graf, and Alexander~V. Balatsky.
\newblock Mapping between finite temperature classical and zero temperature
  quantum systems: Quantum critical jamming and quantum dynamical
  heterogeneities.
\newblock \emph{Phys. Rev. B}, 87:\penalty0 184202, May 2013.
\newblock \doi{10.1103/PhysRevB.87.184202}.
\newblock URL \url{https://link.aps.org/doi/10.1103/PhysRevB.87.184202}.

\bibitem[Bilitewski et~al.(2017)Bilitewski, Zhitomirsky, and
  Moessner]{Bilitewski_2017}
Thomas Bilitewski, Mike~E. Zhitomirsky, and Roderich Moessner.
\newblock Jammed spin liquid in the bond-disordered kagome antiferromagnet.
\newblock \emph{Phys. Rev. Lett.}, 119:\penalty0 247201, Dec 2017.
\newblock \doi{10.1103/PhysRevLett.119.247201}.
\newblock URL \url{https://link.aps.org/doi/10.1103/PhysRevLett.119.247201}.

\bibitem[Bilitewski et~al.(2019)Bilitewski, Zhitomirsky, and
  Moessner]{Bilitewski_2019}
Thomas Bilitewski, Mike~E. Zhitomirsky, and Roderich Moessner.
\newblock Dynamics and energy landscape of the jammed spin liquid.
\newblock \emph{Physical Review B}, 99\penalty0 (5), feb 2019.
\newblock \doi{10.1103/physrevb.99.054416}.
\newblock URL \url{https://doi.org/10.1103%2Fphysrevb.99.054416}.

\bibitem[Silbert et~al.(2005)Silbert, Liu, and Nagel]{SilbertNag2005}
Leonardo~E. Silbert, Andrea~J. Liu, and Sidney~R. Nagel.
\newblock Vibrations and diverging length scales near the unjamming transition.
\newblock \emph{Phys. Rev. Lett.}, 95:\penalty0 098301, Aug 2005.
\newblock \doi{10.1103/PhysRevLett.95.098301}.
\newblock URL \url{https://link.aps.org/doi/10.1103/PhysRevLett.95.098301}.

\bibitem[Vitelli et~al.(2010)Vitelli, Xu, Wyart, Liu, and Nagel]{Nagel09}
Vincenzo Vitelli, Ning Xu, Matthieu Wyart, Andrea~J. Liu, and Sidney~R. Nagel.
\newblock Heat transport in model jammed solids.
\newblock \emph{Phys. Rev. E}, 81:\penalty0 021301, Feb 2010.
\newblock \doi{10.1103/PhysRevE.81.021301}.
\newblock URL \url{https://link.aps.org/doi/10.1103/PhysRevE.81.021301}.

\bibitem[Liarte et~al.(2022)Liarte, Thornton, Schwen, Cohen, Chowdhury, and
  Sethna]{DL21a}
Danilo~B. Liarte, Stephen~J. Thornton, Eric Schwen, Itai Cohen, Debanjan
  Chowdhury, and James~P. Sethna.
\newblock Universal scaling for disordered viscoelastic matter near the onset
  of rigidity.
\newblock \emph{Phys. Rev. E}, 106:\penalty0 L052601, Nov 2022.
\newblock \doi{10.1103/PhysRevE.106.L052601}.
\newblock URL \url{https://link.aps.org/doi/10.1103/PhysRevE.106.L052601}.

\bibitem[{Liarte} et~al.(2022){Liarte}, {Thornton}, {Schwen}, {Cohen},
  {Chowdhury}, and {Sethna}]{DL21b}
Danilo~B. {Liarte}, Stephen~J. {Thornton}, Eric {Schwen}, Itai {Cohen},
  Debanjan {Chowdhury}, and James~P. {Sethna}.
\newblock {Universal scaling for disordered viscoelastic matter II: Collapses,
  global behavior and spatio-temporal properties}.
\newblock \emph{arXiv e-prints}, art. arXiv:2202.13933, February 2022.

\bibitem[Thornton et~al.(2023)Thornton, Liarte, Abbamonte, Sethna, and
  Chowdhury]{SuppMaterial}
S.~Thornton, D.B. Liarte, P.~Abbamonte, J.P. Sethna, and D.~Chowdhury.
\newblock Supplementary information: Jamming and unusual charge density
  fluctuations of strange metals, 2023.

\bibitem[DeGiuli et~al.(2014)DeGiuli, Laversanne-Finot, Düring, Lerner, and
  Wyart]{GiuliWya2014}
Eric DeGiuli, Adrien Laversanne-Finot, Gustavo Düring, Edan Lerner, and
  Matthieu Wyart.
\newblock Effects of coordination and pressure on sound attenuation{,} boson
  peak and elasticity in amorphous solids.
\newblock \emph{Soft Matter}, 10:\penalty0 5628--5644, 2014.
\newblock \doi{10.1039/C4SM00561A}.
\newblock URL \url{http://dx.doi.org/10.1039/C4SM00561A}.

\bibitem[Franz et~al.(2015)Franz, Parisi, Urbani, and Zamponi]{FranzZam2015}
Silvio Franz, Giorgio Parisi, Pierfrancesco Urbani, and Francesco Zamponi.
\newblock Universal spectrum of normal modes in low-temperature glasses.
\newblock \emph{Proceedings of the National Academy of Sciences}, 112\penalty0
  (47):\penalty0 14539--14544, 2015.
\newblock ISSN 0027-8424.
\newblock \doi{10.1073/pnas.1511134112}.
\newblock URL \url{https://www.pnas.org/content/112/47/14539}.

\bibitem[Kohsaka et~al.(2007)Kohsaka, Taylor, Fujita, Schmidt, Lupien,
  Hanaguri, Azuma, Takano, Eisaki, Takagi, Uchida, and Davis]{disorder1}
Y.~Kohsaka, C.~Taylor, K.~Fujita, A.~Schmidt, C.~Lupien, T.~Hanaguri, M.~Azuma,
  M.~Takano, H.~Eisaki, H.~Takagi, S.~Uchida, and J.~C. Davis.
\newblock An intrinsic bond-centered electronic glass with unidirectional
  domains in underdoped cuprates.
\newblock \emph{Science}, 315\penalty0 (5817):\penalty0 1380--1385, 2007.
\newblock \doi{10.1126/science.1138584}.
\newblock URL \url{https://www.science.org/doi/abs/10.1126/science.1138584}.

\bibitem[Pan et~al.(2001)Pan, O'Neal, Badzey, Chamon, Ding, Engelbrecht, Wang,
  Eisaki, Uchida, Gupta, Ng, Hudson, Lang, and Davis]{disorder2}
S.~H. Pan, J.~P. O'Neal, R.~L. Badzey, C.~Chamon, H.~Ding, J.~R. Engelbrecht,
  Z.~Wang, H.~Eisaki, S.~Uchida, A.~K. Gupta, K.-W. Ng, E.~W. Hudson, K.~M.
  Lang, and J.~C. Davis.
\newblock Microscopic electronic inhomogeneity in the high-{Tc} superconductor
  {Bi2Sr2CaCu2O8}+x.
\newblock \emph{Nature}, 413\penalty0 (6853):\penalty0 282--285, September
  2001.
\newblock ISSN 1476-4687.
\newblock \doi{10.1038/35095012}.
\newblock URL \url{https://doi.org/10.1038/35095012}.

\bibitem[Olson~Reichhardt et~al.(2012)Olson~Reichhardt, Groopman, Nussinov, and
  Reichhardt]{ReichhardtGNR12}
C.~J. Olson~Reichhardt, E.~Groopman, Z.~Nussinov, and C.~Reichhardt.
\newblock Jamming in systems with quenched disorder.
\newblock \emph{Phys. Rev. E}, 86:\penalty0 061301, Dec 2012.
\newblock \doi{10.1103/PhysRevE.86.061301}.
\newblock URL \url{https://link.aps.org/doi/10.1103/PhysRevE.86.061301}.

\bibitem[{Graves} et~al.(2016){Graves}, {Nashed}, {Padgett}, {Goodrich}, {Liu},
  and {Sethna}]{GravesNPGLS16}
A.~L. {Graves}, S.~{Nashed}, E.~{Padgett}, C.~P. {Goodrich}, A.~J. {Liu}, and
  J.~P. {Sethna}.
\newblock Pinning susceptibility: The effect of dilute, quenched disorder on
  jamming.
\newblock \emph{Phys. Rev. Lett.}, 116, June 2016.

\bibitem[P{\'e}ter et~al.(2018)P{\'e}ter, Lib{\'a}l, Reichhardt, and
  Reichhardt]{peter2018crossover}
Huba P{\'e}ter, Andr{\'a}s Lib{\'a}l, Charles Reichhardt, and CJ~Olson
  Reichhardt.
\newblock Crossover from jamming to clogging behaviours in heterogeneous
  environments.
\newblock \emph{Scientific reports}, 8\penalty0 (1):\penalty0 1--9, 2018.

\bibitem[Zhang et~al.(2022)Zhang, Ridout, Parts, Sachdeva, Bester,
  Vollmayr-Lee, Utter, Brzinski, and Graves]{ZhangRPSBVUBG22}
Andy~L. Zhang, Sean~A. Ridout, Celia Parts, Aarushi Sachdeva, Cacey~S. Bester,
  Katharina Vollmayr-Lee, Brian~C. Utter, Ted Brzinski, and Amy~L. Graves.
\newblock Jammed solids with pins: Thresholds, force networks, and elasticity.
\newblock \emph{Phys. Rev. E}, 106:\penalty0 034902, Sep 2022.
\newblock \doi{10.1103/PhysRevE.106.034902}.
\newblock URL \url{https://link.aps.org/doi/10.1103/PhysRevE.106.034902}.

\bibitem[Schmalian and Wolynes(2000)]{schmalian}
J\"org Schmalian and Peter~G. Wolynes.
\newblock Stripe glasses: Self-generated randomness in a uniformly frustrated
  system.
\newblock \emph{Phys. Rev. Lett.}, 85:\penalty0 836--839, Jul 2000.
\newblock \doi{10.1103/PhysRevLett.85.836}.
\newblock URL \url{https://link.aps.org/doi/10.1103/PhysRevLett.85.836}.

\bibitem[Facoetti et~al.(2019)Facoetti, Biroli, Kurchan, and Reichman]{kurchan}
Davide Facoetti, Giulio Biroli, Jorge Kurchan, and David~R. Reichman.
\newblock Classical glasses, black holes, and strange quantum liquids.
\newblock \emph{Phys. Rev. B}, 100:\penalty0 205108, Nov 2019.
\newblock \doi{10.1103/PhysRevB.100.205108}.
\newblock URL \url{https://link.aps.org/doi/10.1103/PhysRevB.100.205108}.

\bibitem[Bouzid et~al.(2018)Bouzid, Keshavarz, Geri, Divoux, Del~Gado, and
  McKinley]{EDGSoftGels}
Mehdi Bouzid, Bavand Keshavarz, Michela Geri, Thibaut Divoux, Emanuela
  Del~Gado, and Gareth~H. McKinley.
\newblock Computing the linear viscoelastic properties of soft gels using an
  optimally windowed chirp protocol.
\newblock \emph{Journal of Rheology}, 62\penalty0 (4):\penalty0 1037--1050,
  2018.
\newblock \doi{10.1122/1.5018715}.
\newblock URL \url{https://doi.org/10.1122/1.5018715}.

\bibitem[Colombo and Del~Gado(2014)]{EDGColloidGel}
Jader Colombo and Emanuela Del~Gado.
\newblock Stress localization, stiffening, and yielding in a model colloidal
  gel.
\newblock \emph{Journal of Rheology}, 58\penalty0 (5):\penalty0 1089--1116,
  2014.
\newblock \doi{10.1122/1.4882021}.
\newblock URL \url{https://doi.org/10.1122/1.4882021}.

\bibitem[Kohn and Milton(1986)]{KohnAveraging}
Robert~V. Kohn and Graeme~W. Milton.
\newblock On bounding the effective conductivity of anisotropic composites.
\newblock In J.~L. Ericksen, David Kinderlehrer, Robert Kohn, and J.-L. Lions,
  editors, \emph{Homogenization and Effective Moduli of Materials and Media},
  pages 97--125, New York, NY, 1986. Springer New York.
\newblock ISBN 978-1-4613-8646-9.

\end{thebibliography}

\setcounter{figure}{0}
\renewcommand\thefigure{S\arabic{figure}}
\renewcommand{\theHfigure}{S\thefigure}

\clearpage
\pagebreak
\onecolumngrid

\begin{center}
\Large\textbf{Supplementary Information}
\end{center}
\section{Effective medium theory}
\begin{figure*}[!ht]
\begin{center}
\includegraphics[width=0.65\linewidth]{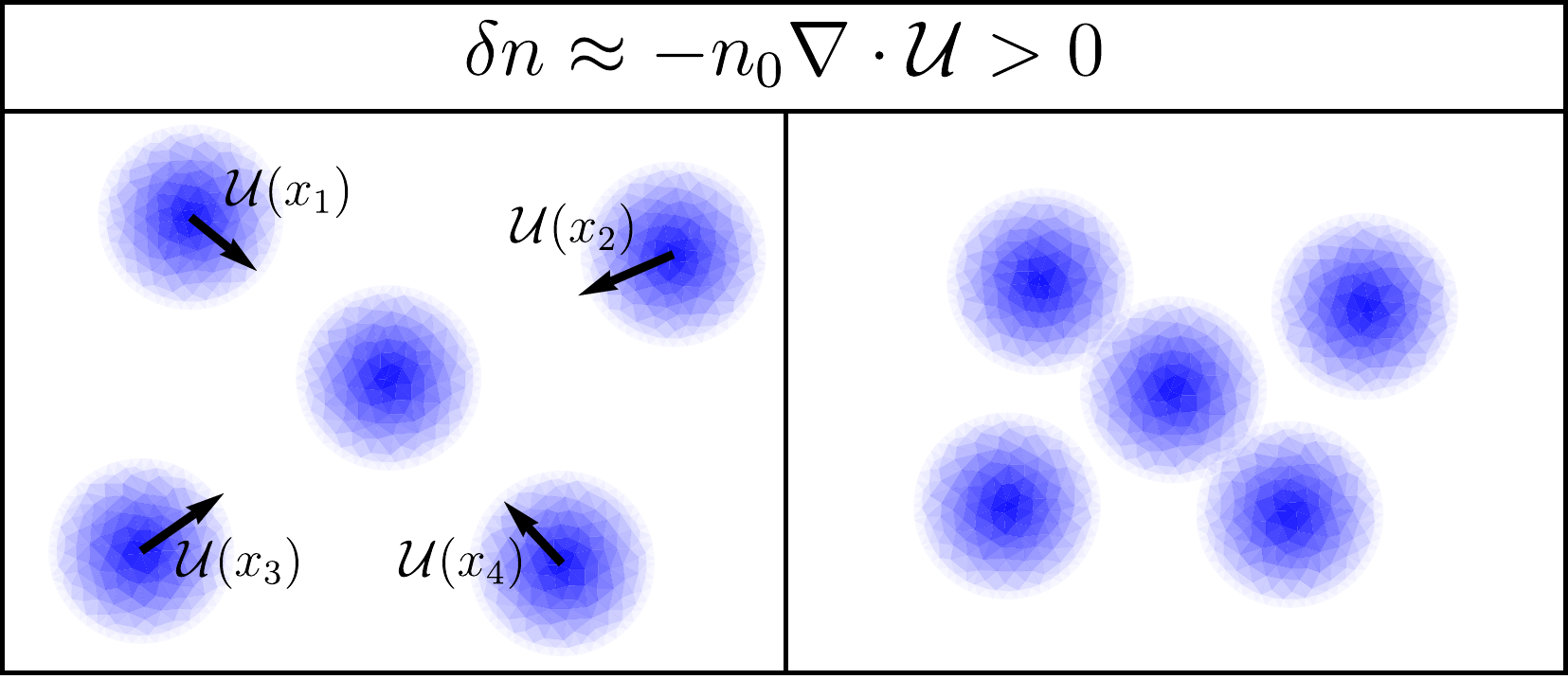}
\end{center}
\caption{The density-density response is connected to longitudinal fluctuations of the displacement. Shown here is a local compression wave increasing the local density of charged particles; the precise composition of the charged low-energy excitations is not known.
\label{fig:DensityResponse}}
\end{figure*}

As noted in the main text, we model the system as a layered structure consisting of two-dimensional randomly populated lattices of harmonic springs. The local deformations associated with the configurations (e.g. compression) are assumed to capture the fluctuations associated with an underlying electronic liquid; see Fig.~\ref{fig:DensityResponse}. We make an analogy between the strongly overdamped plasmon in a class of strange metals and the strongly overdamped phonon in viscoelastic systems near the onset of rigidity. One of the features of rigidity percolation (and jamming) is a nearly flat density of states at the transition, which corresponds to an anomalously large number of low-energy excitations into which an excited phonon can decay. This is captured in randomly populated networks analytically by an effective-medium theory known as the coherent potential approximation (CPA) \cite{LiarteLub2019}.

In this model, we begin with a fully occupied lattice with unit elastic coefficients (harmonic springs) and randomly dilute the lattice, keeping each bond with probability $p$. To study the dynamics of the disordered system, the CPA self-consistently replaces the Green's function of this randomly populated lattice with a spatially homogeneous disorder-averaged Green's function. The self-energy calculated in this framework is typically recast as a frequency-dependent elastic coefficient $\mathcal{K}\left(\omega\right)$. More generally, given several bond occupation probabilities $\left\{p_{\alpha}\right\}$, the CPA gives frequency-dependent elastic coefficients for each sublattice $\left\{\mathcal{K}_{\alpha}\left(\omega\right)\right\}$, where $\alpha=1,2\dots$ is the sublattice index. This method faithfully reproduces the zero-frequency athermal phase diagram and phase transitions predicted by Maxwell's constraint counting arguments, where the number of constraints is equal to the number of degrees of freedom \cite{LiarteLub2019}. There is also experimental evidence that it properly reproduces the low-frequency behavior of other microscopically disordered systems undergoing rigidity transitions, such as soft gels \cite{EDGSoftGels, EDGColloidGel}.

A specific type of phase transition that these randomly populated lattices can undergo is broadly known as \textit{rigidity percolation}, when a disordered elastic network loses rigidity in a continuous fashion at $p=p_c$. This can be seen by examining the effective bulk modulus $B$ and the shear modulus $G$ of a bond-diluted lattice, defined by how the potential energy of the lattice $U$ changes in response to small linearized strains $u_{ij}$:
\beq
\Delta U=\frac{B}{2}\left(u_{xx}+u_{yy}\right)^2+2G\left[u_{xy}^2+\frac{1}{4}\left(u_{xx}-u_{yy}\right)^2\right].
\eeq
The predicted zero-frequency bulk and shear moduli of these diluted lattices grow continuously from $0$ across such a transition, as $p$ grows larger than $p_c$. The density-density response near jamming is discontinuous across the critical point due to the jump in the bulk modulus, as opposed to the response near RP. The MEELS phenomenology is better captured in terms of RP, even though the nature of the anomalous low-energy excitations in the vibrational density of states is the same for both. This paper focuses on the \textit{universal} behavior of \textit{rigidity percolation}, that is, the characteristics of the phase transition that do not depend on the underlying microscopic lattice geometry.

For a single sublattice in two dimensions, the self-consistent equation for the elastic coefficients reads
\beq
\frac{p-\mathcal{K}\left(\omega\right)}{1-\mathcal{K}\left(\omega\right)}=\frac{1}{z}\frac{1}{s_{\textsc{BZ}}}\int_{\textsc{BZ}}\textrm{d}^2q\;\textrm{Tr}\left[D\left(q,\mathcal{K}\left(\omega\right)\right)\left(D\left(q,\mathcal{K}\left(\omega\right)\right)-\omega^2I\right)^{-1}\right]
\eeq
where $p$ is as defined earlier, $D(...)$ is the dynamical matrix, and $z$ is the average number of bonds per unit cell. The derivation of the more general form of this expression can be found, for instance, in the supplemental material of \cite{LiarteLub2019}.

By expanding the dynamical matrix for a general 2D isotropic lattice  in the long-wavelength limit as
\beq
D_{ij}\left(\mathcal{K}\left(\omega\right)\right)=B\left(\mathcal{K}\left(\omega\right)\right)q^2\hat{q}_i\hat{q}_j+G\left(\mathcal{K}\left(\omega\right)\right)q^2\delta_{ij},
\eeq
we can evaluate the integral and solve the self-consistency equations. The combination $K_{\textsc{L}}\equiv B+G$ is the only combination of moduli that enters the longitudinal response (and the density-density response), and must be proportional to the only microscopic coupling in the problem $\mathcal{K}$. Writing the self-consistent equation for $K_{\textsc{L}}$ and expanding to linear order in the distance to the continuous critical point $\DpRP\equiv p-p_c$, one finds an expression involving only certain invariant scaling combinations. The coupling is found as the self-consistent solution to
\begin{subequations}
\begin{align}
\pm1-\widetilde{K}_{\textsc{L}\pm} &= \frac{\widetilde{\omega}'^2}{\widetilde{K}_{\textsc{L}\pm}}\log\left(-\frac{\widetilde{K}_{\textsc{L}\pm}}{\widetilde{\omega}'^2\widetilde{Z}}\right),\\
\widetilde{K}_{\textsc{L}\pm}&\equiv\frac{K_{\textsc{L}\pm}}{K_{\textsc{L}0}\left|\DpRP\right|},\\
\widetilde{\omega}'&\equiv\frac{\omega}{\omega'_0\left|\DpRP\right|},\\
\widetilde{Z}&\equiv Z_0\left|\DpRP\right|.
\end{align}
\end{subequations}
Here $K_{\textsc{L}0}$, $\omega'_0$, and $Z_0$ are non-universal constants of order unity, and we choose their values from the bond-diluted mechanical triangular lattice with nearest-neighbor bonds for all plots. The $+$ and $-$ in the self-consistent equations refer to the ``solid'' ($p>p_c$) and ``floppy'' ($p<p_c$) sides of the transition, respectively. Asymptotically close to the transition, the solution to the above self-consistent effective medium equation is given by
\beq
\widetilde{K}'_{\textsc{L}\pm}\left(\widetilde{\omega}''\right)=\pm1+\sqrt{1-\widetilde{\omega}''^2},
\eeq
with $\widetilde{K}'_{\textsc{L}\pm}\equiv K_{\textsc{L}\pm}/K''_{\textsc{L}0}\left|\DpRP\right|$ and $\widetilde{\omega}''\equiv\omega/\omega_0''\left(\left|\DpRP\right|/\left|\log\left(\left|\DpRP\right|\right)\right|^{1/2}\right)$. The logarithm that appears in this scaling variable for the frequency is unique to $2$ dimensions and does not affect the qualitative results of any of our calculations; this will be elaborated upon in a future manuscript. The imaginary part of the square root should be interpreted as non-positive for causality reasons. The dissipation that is calculated within the CPA framework aims to capture phonon scattering off of ``defects'' introduced by the disorder. The vanishing of the imaginary part of $K_{\textsc{L}}\left(\omega\right)$ at small but finite frequency does not survive universal corrections to scaling; when the full CPA self-consistency equation is solved on the solid side ($\DpRP>0$), then $\textrm{Im}\left(K_{\textsc{L}}\right)<0$ for all $\omega>0$.
\section{Universal scaling of the density-density response}
The frequency-dependent elastic coefficient derived in the previous section can be used to determine the effective bulk and shear moduli of the elastic medium, which can be used to determine the long-wavelength density-density response. The form of our long-wavelength density-density response is calculated as follows: the equation of motion for an isotropic elastic sheet without external damping reads
\beq
\rho\,\ddot{\mathbf{\mathcal{U}}}=B\nabla\left(\nabla\cdot\mathbf{\mathcal{U}}\right)+G\nabla^2\mathbf{\mathcal{U}}+\mathbf{f}^{\textrm{ext}}
\eeq 
where $\rho$ is the (constant) average background density and $\mathcal{U}$ is a small displacement field. $B$ and $G$ are proportional to $\mathcal{K}\left(\omega\right)$, and are hence frequency-dependent and complex. Assuming the local perturbation due to the external forcing field to be small, we expand $n\approx n_0\left(1-\nabla\cdot\mathcal{U}\right)=\rho/m\left(1-\nabla\cdot\mathcal{U}\right)$; see Fig.~\ref{fig:DensityResponse}. 

We then use the definition of the susceptibility as the change in the density as a result of the perturbing conjugate field, $\Pi\equiv-\delta n/\delta h$. Note that the additional negative sign in front of the susceptibility compared to \cite{DL21a,DL21b} is related to differing definitions for the susceptibility; we adopt the experimentalists' convention \cite{Abbamonte1} where the imaginary part is negative for positive frequencies. Just as $\mathcal{U}$ and $f$ are thermodynamic conjugates (because they enter the energy density as $\mathcal{U}\cdot f$), so are $\delta n_q=in_0q\cdot\mathcal{U}_q$ and $\delta h_q=f_q^{\textsc{L}\textrm{ext}}/in_0q$. Thus,
\beq
\left(-\rho\omega^2+\left(B+G\right)q^2\right)\delta n_q=n_0^2q^2\delta h_q,\\
\Pi\equiv-\frac{\delta n_q}{\delta h_q}=\frac{\rho^2q^2}{m^2}\frac{1}{\rho\omega^2-K_L\left(\omega\right)q^2}.
\eeq
This allows us to write a universal form for the density-density response at rigidity percolation on both sides of the transition:
\beq
\widetilde{\Pi}=\frac{\widetilde{q}^2}{\widetilde{\omega}''^2-\left(\sqrt{1-\widetilde{\omega}''^2}\pm1\right)\widetilde{q}^2}
\eeq
where the appropriate scaling is found to be $\widetilde{q}\equiv q/q_0\left(\left|\DpRP\right|^{1/2}/\left|\log\left(\left|\DpRP\right|\right)\right|^{1/2}\right)$ and $\widetilde{\Pi}\equiv\Pi\left|\DpRP\right|/\Pi_0$. The $O(1)$ constants $q_0$ and $\Pi_0$ are non-universal. Since the scaling of $B$, $G$, and $\omega$ near rigidity percolation has already been fixed by the self-consistency equation, the scaling of $q$ is found by balancing powers of $\delta p$ in the denominator of the expression for $\Pi$. In practice, the corrections to scaling that fix the imaginary part at low frequencies are significant enough (they vanish as $\sim\left|\log\left|\delta p\right|\right|^{-1/2}$) that we numerically solve the self-consistent equations with $\widetilde{Z}$ included to find a more faithful representation of the CPA predictions. Finally, the effects of the long-ranged Coulomb interaction are added using the RPA (as described in the main text):
\beq
\chi(q,\omega) = \frac{\Pi(q,\omega)}{1-V(q)\Pi(q,\omega)} 
\eeq
with the 3D Coulomb interaction $V(q)=4\pi e^2/q^2$.

The qualitative features that the CPA predicts for $\chi$ reflect what is expected of a lattice near a rigidity transition: sharply defined quasiparticles exist only at the longest wavelengths, and rapidly broaden with increasing $q$ into an incoherent bump at a frequency $\Delta\omega_{\star}$ set by the distance to the rigidity percolation transition. If we are near the critical point, experimental probes of the response will inevitably probe only the region of large $\widetilde{q}$, which leads to a $q$-independent shape of the response. In this model, a coherent quasiparticle can still be found near the center of the BZ.

For large values of $\widetilde{q}$, there is a range of frequencies where $\chi''\sim\omega^{-1}$ decays slowly. For an experimental probe, this may indicate the violation of \textit{f}-sum rules. However, at frequencies large enough, $\chi''$ eventually decays as $\sim\omega^{-3}$ as predicted by Drude theory. Although the paradigm of proximity to a critical point with a large number of anomalously low-frequency modes serves to explain the $q$-independent shape of the response outside of the very center of the BZ, the shapes of the universal forms of the electronic response do not have the plateaus measured by the experiment. Because we have control over how our response depends on the distance to the critical point, we can investigate how long-wavelength disorder in a sample, represented by a distribution of distances to the critical point $\DpRP$, modifies the observed form of the density-density response.
\section{Averaging over the long-wavelength sample disorder}
We imagine an experimental sample prepared on average close to a critical point $\DpRP$, where different regions of the sample are allowed to have slightly different distances to the critical point with spread $\sigma$. The effect of this long-wavelength disorder is then represented by averaging the response over many distances to the critical point using a Gaussian of center $\DpRP$ and width $\sigma$: 
\beq
\overline{\chi_{\DpRP}(q,\omega)} &=& 
\int_{-\infty}^{\infty}d\left(\D\pRP'\right)  P_\sigma(\D\pRP')~ \chi_{\DpRP'}(q,\omega)\\
P_\sigma[\D\pRP'] &=&\frac{1}{\sqrt{2\pi\sigma^2}}e^{-\left(\D\pRP'\right)^2/2\sigma^2},
\eeq
where $\chi_{\DpRP}(q,\omega)$ is the response at a fixed distance ($\DpRP$) from RP, and we choose a Gaussian distribution, $P_\sigma[\D\pRP]$, with width $\sigma$ and $\D\pRP'\equiv\DpRP'-\DpRP$. One could imagine other methods for averaging over the effect of the long-wavelength disorder. In this case, we adopt the procedure where the susceptibility (related to an inverse stiffness at zero frequency) is estimated as an arithmetic mean of the susceptibilities of different portions of the sample; this means that a large susceptibility in any portion of the sample corresponds to a large average susceptibility. One could also estimate the susceptibility as the inverse of the average inverse susceptibility, which would suppress the susceptibility if there is a small susceptibility in any portion of the sample. For an inhomogeneous medium, the effective susceptibility is bounded between these two options \cite{KohnAveraging}.

\begin{figure*}[!ht]
\begin{center}
\begin{minipage}{0.22\linewidth}    
\hspace{0.25cm} \vspace{0.2cm} \\
    (a) \vspace{0.1cm} \\
    \includegraphics[height=0.95\linewidth]{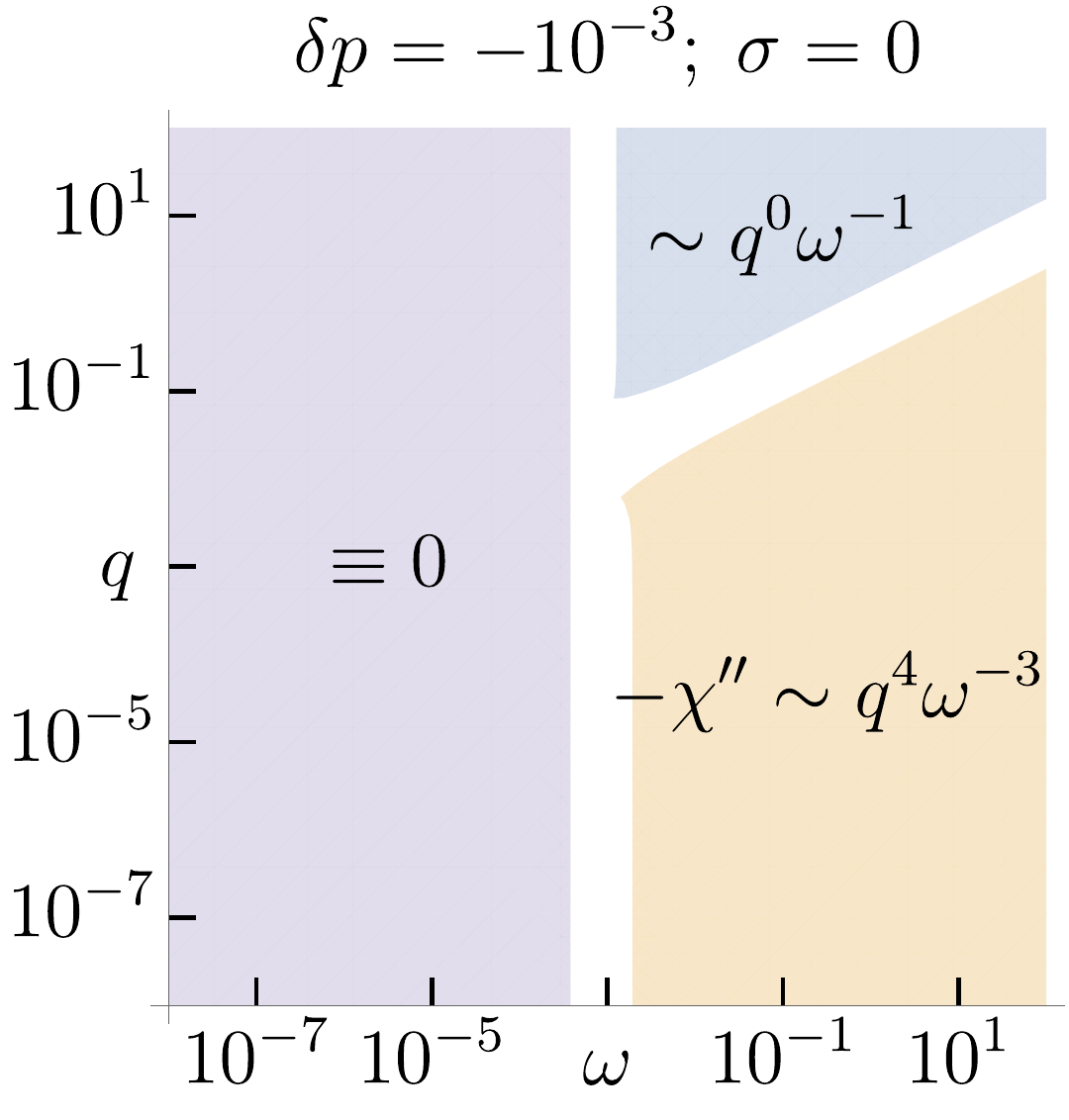}
\end{minipage}\hspace{0.03\linewidth}
\begin{minipage}{0.22\linewidth}
\hspace{0.25cm} \vspace{0.2cm} \\
    (b) \vspace{0.1cm} \\
    \includegraphics[height=0.95\linewidth]{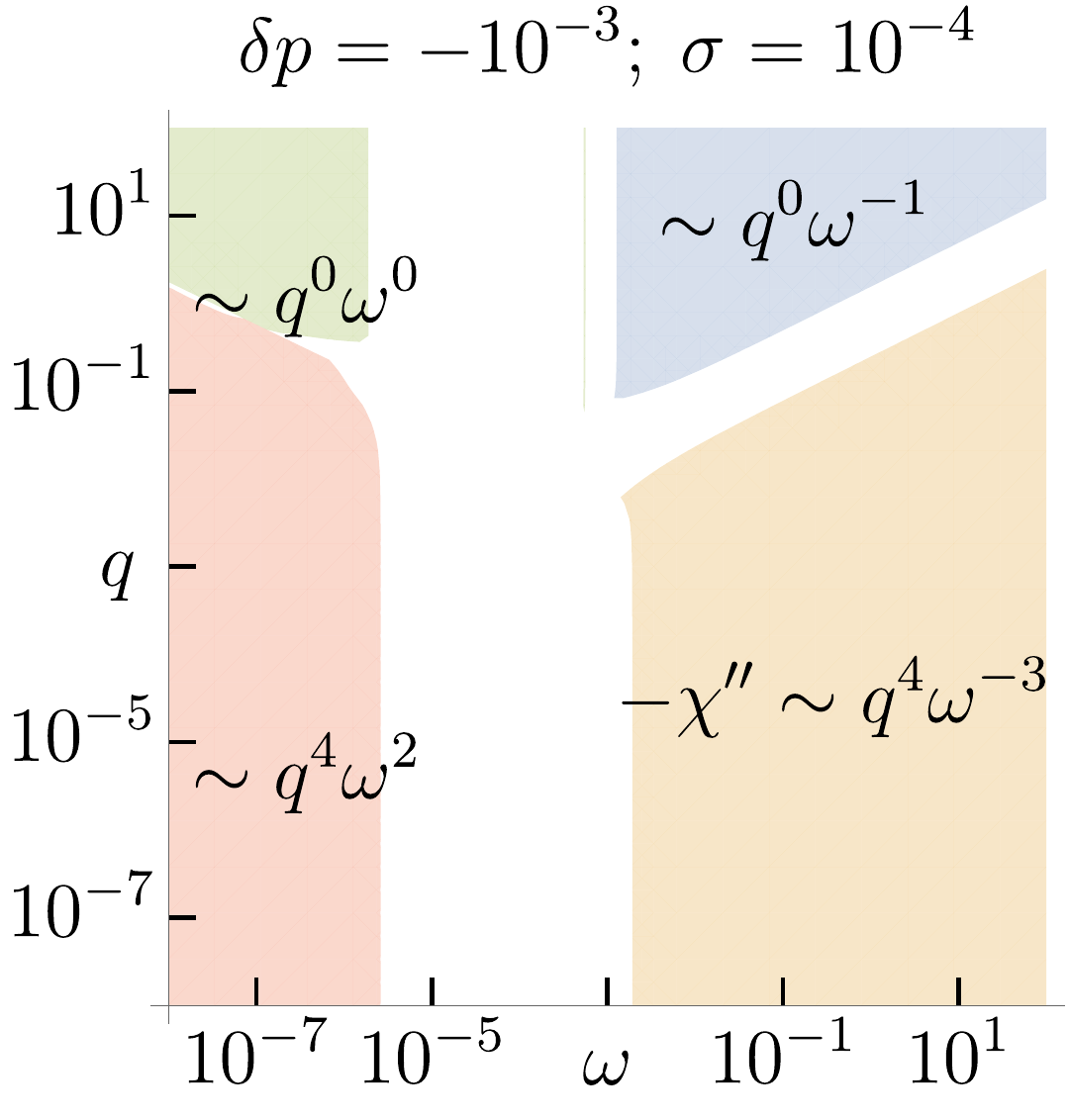}
\end{minipage}\hspace{0.03\linewidth}
\begin{minipage}{0.22\linewidth}
\hspace{0.25cm} \vspace{0.2cm} \\
    (c) \vspace{0.1cm} \\
    \includegraphics[height=0.95\linewidth]{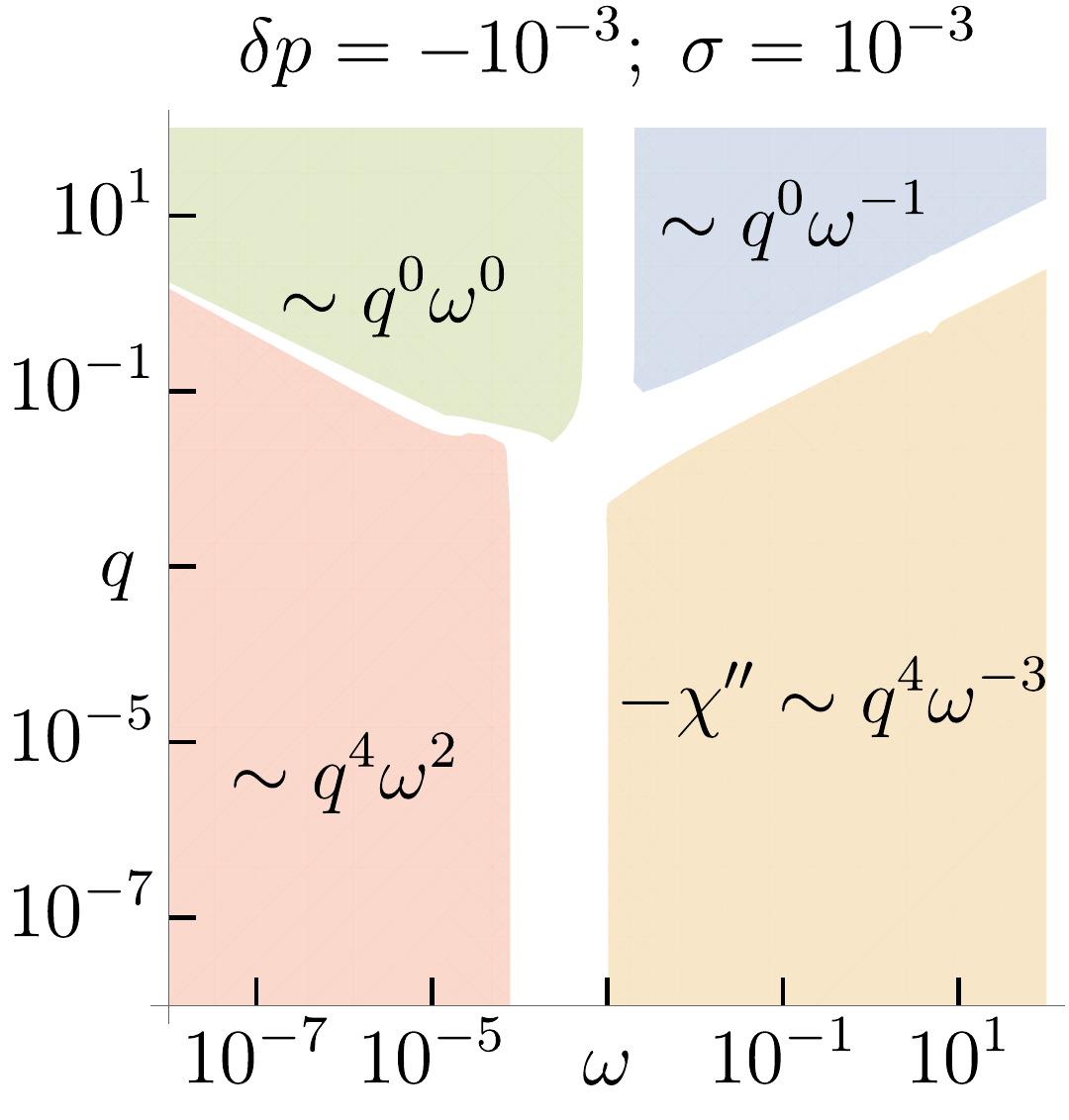}
\end{minipage}\hspace{0.03\linewidth}
\begin{minipage}{0.22\linewidth}
\hspace{0.25cm} \vspace{0.2cm} \\
    (d) \vspace{0.1cm} \\
    \includegraphics[height=0.95\linewidth]{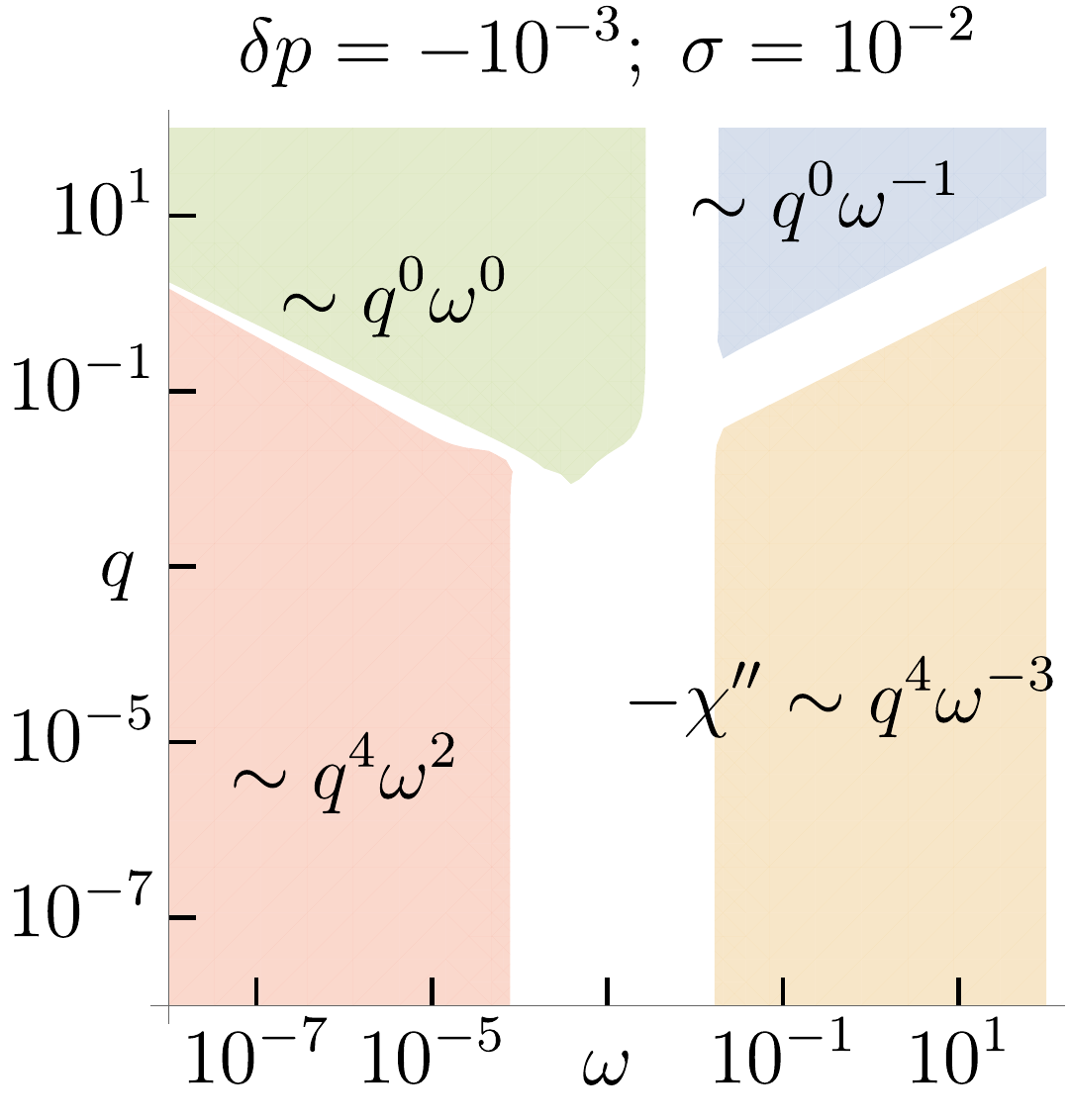}
\end{minipage}
\vspace{0.1 cm} \\
\begin{minipage}{0.22\linewidth}
    (e) \vspace{0.1cm} \\
    \includegraphics[height=0.95\linewidth]{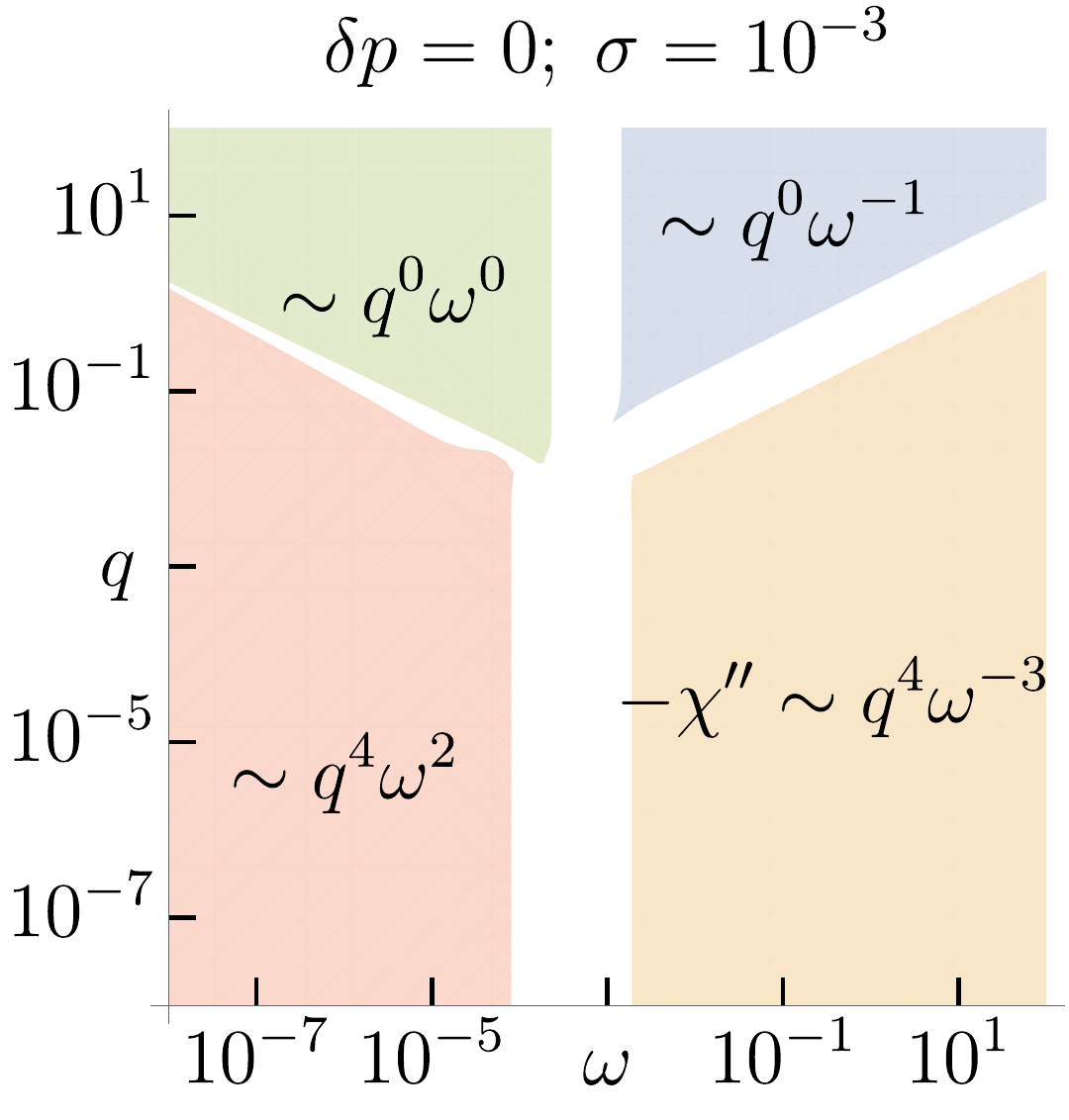}
\end{minipage}\hspace{0.03\linewidth}
\begin{minipage}{0.22\linewidth}
    (f) \vspace{0.1cm} \\
    \includegraphics[height=0.95\linewidth]{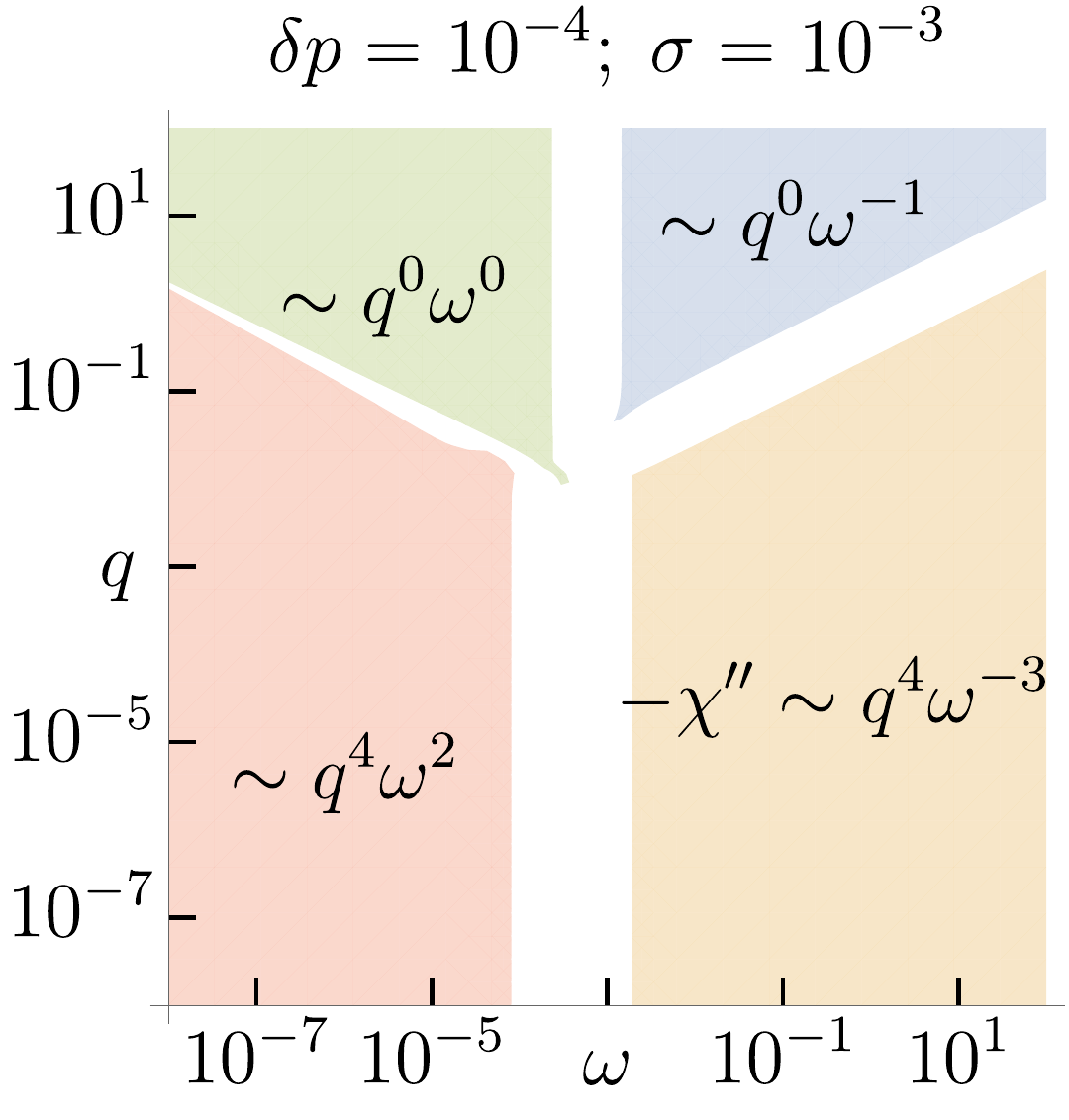}
\end{minipage}\hspace{0.03\linewidth}
\begin{minipage}{0.22\linewidth}
    (g) \vspace{0.1cm} \\
    \includegraphics[height=0.95\linewidth]{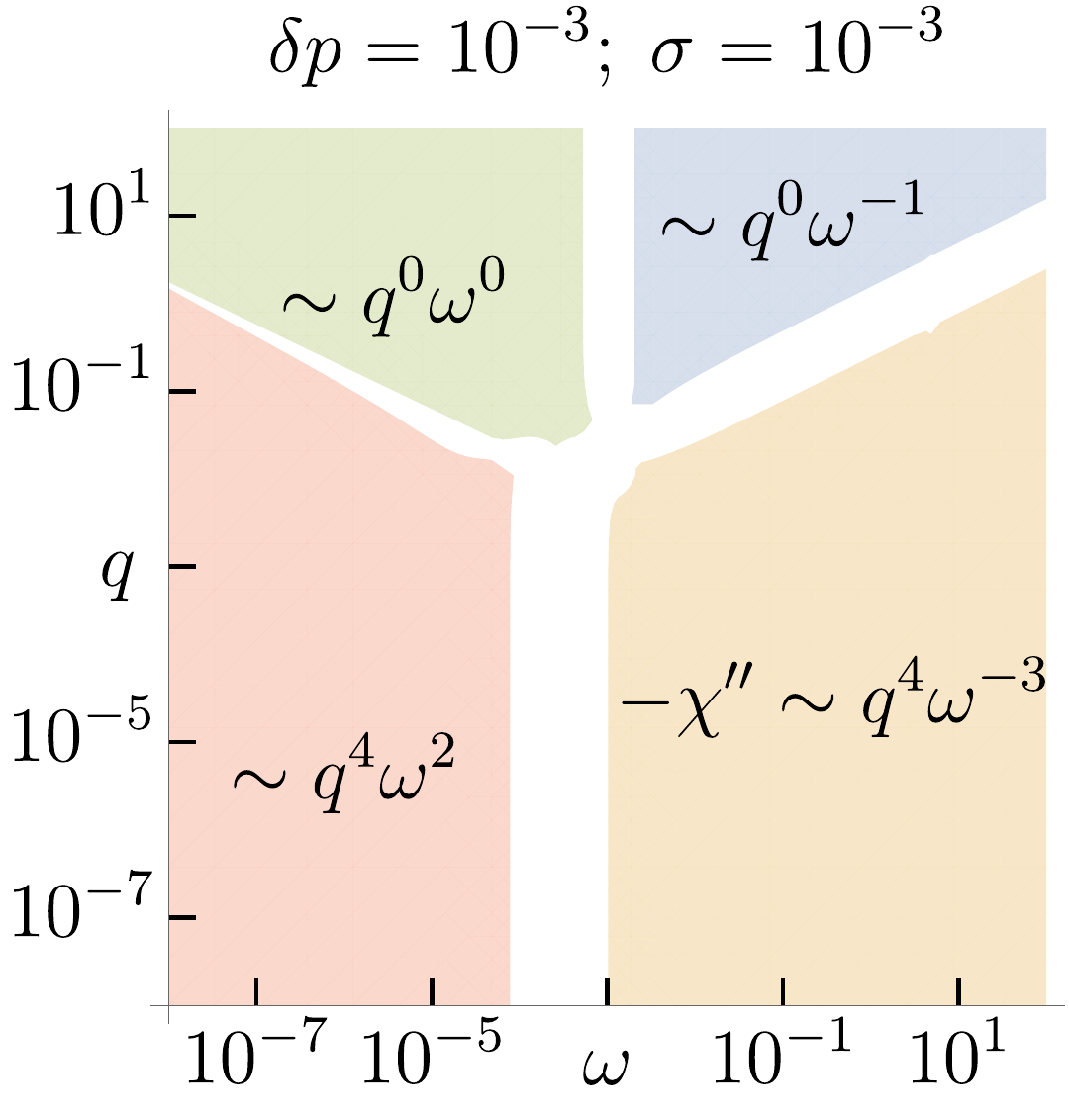}
\end{minipage}\hspace{0.03\linewidth}
\begin{minipage}{0.22\linewidth}
    (h) \vspace{0.1cm} \\
    \includegraphics[height=0.95\linewidth]{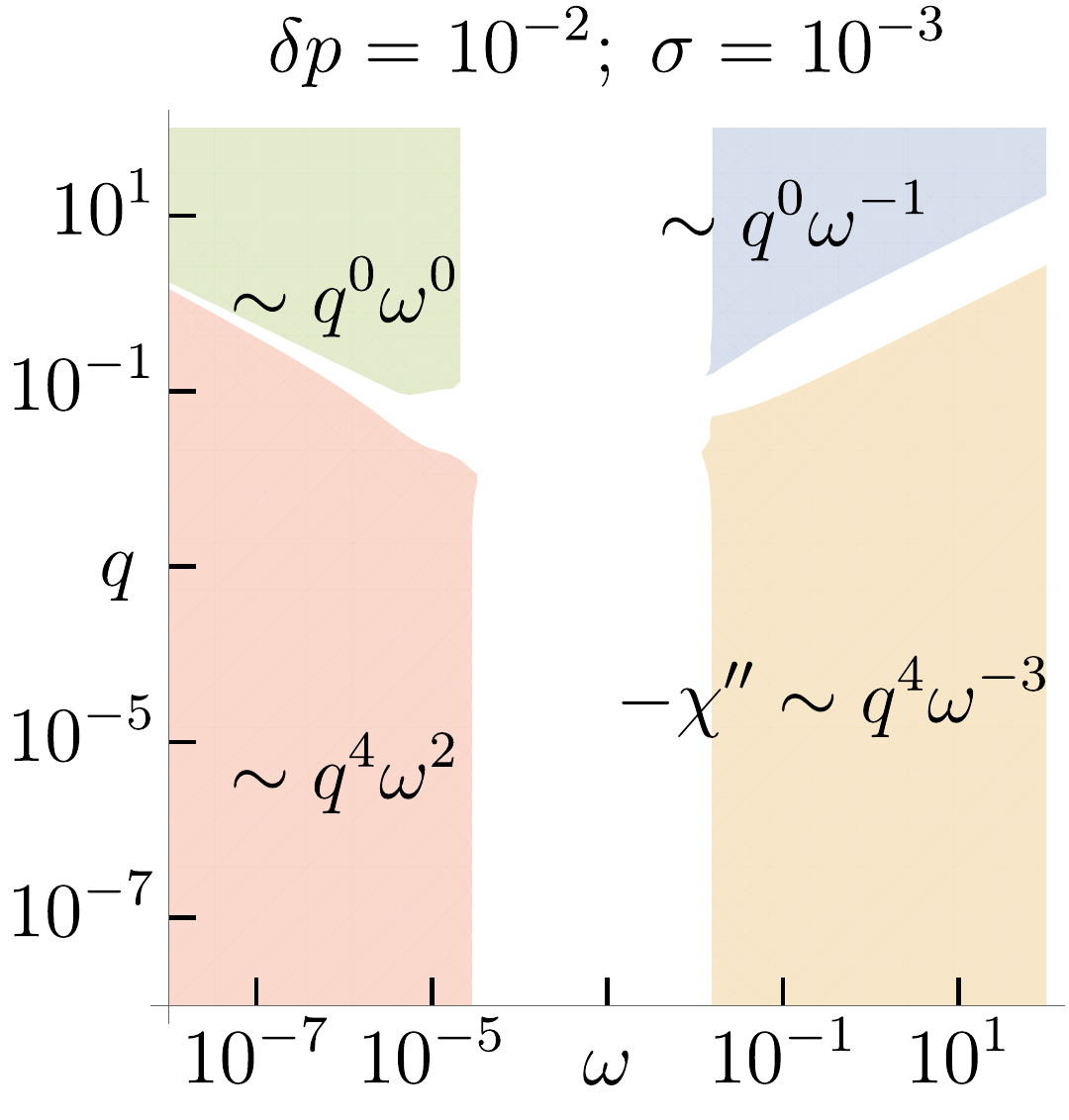}
\end{minipage}
\end{center}
\caption{Scaling regimes of the disorder-averaged susceptibilities near the rigidity transition. Plots (a)--(d) show fixed $\delta p=-10^{-3}$ near the transition on the liquid side with increasing $\sigma$, demonstrating the emergence of the disorder-averaged plateau. Plots (e)--(h) show the effect of varying $\delta p$ into the solid side at fixed $\sigma=10^{-3}$. In this case, the disorder-averaging becomes less relevant and the peak shifts into higher frequencies with the predicted scaling. The region of exactly $0$ imaginary part in (a) is eliminated by corrections to scaling. Plots are similar at large $q$ on the solid and liquid sides. 
\label{fig:Regions}}
\end{figure*}

The solution to the full self-consistency equations including the corrections to scaling cannot be written in terms of elementary functions, so we estimate the convolution by picking many values of $\DpRP$, computing $\chi''$ over a range of frequencies by solving the full self-consistency equations, and performing an appropriate weighted average of the $\chi''$ at the experimental values of $q$ (which correspond to asymptotically large $\widetilde{q}$ for windows of $\DpRP$ very close to the critical point). These averaged responses $\overline{\chi}''$ are compared with the measured imaginary part of the density-density response from the experiment. We seek to describe the universal behaviour of the electronic response, and so we will not capture the additional \textit{lattice} phonon seen in the experimental data at the lowest frequencies in \cite{Abbamonte2}. The plots in Fig.~\ref{fig:Regions} are generated near rigidity percolation using a width of $\sigma=1.2\times 10^{-3}$ in probability space. The central distances to RP chosen are $\DpRP=1.0\times 10^{-3}$ and $\DpRP=1.4\times 10^{-3}$ and exhibit $q$-independence over a selected range of frequencies for $q\gg \left|\delta p\right|^{1/2}/\left|\log\left(\left|\delta p\right|\right)\right|^{1/2}$.

Without disorder averaging, for $q\gg \left|\delta p\right|^{1/2}/\left|\log\left(\left|\delta p\right|\right)\right|^{1/2}$, the susceptibility has a broad bump near $\omega_{\star}$ associated with the decay of the quasiparticle into the anomalous low-frequency modes. When disorder averaging is performed, a sequence of these bumps leads to the emergence of a low-frequency plateau terminating near $\omega_{\star}$. Above the plateau, this theory predicts a region where $\chi''\sim\omega^{-1}$ crossing over into a region where $\chi''\sim\omega^{-3}$ (the high-frequency Drude scaling). In the experiment, an exponent $1\leq\alpha\leq3$ is measured out to the highest frequencies.

The imaginary part of the susceptibility $\chi''$ is odd in $\omega$ but contains a plateau that extends down to the lowest frequencies with increasing $q$. This results in a sharp feature at the lowest frequencies where the plateau ends and the response crosses over into power-law behavior to ensure $\chi''(q,0)=0$. The location of this feature can be deduced from Fig.~\ref{fig:Regions}, where the green ($-\chi''\sim q^0\omega^0$) and red ($-\chi''\sim q^4\omega^2$) regions touch.  Plots of the lowest frequency behavior are shown below in Fig.~\ref{fig:sharpfeat} for the sake of completeness. In the experiment, the low frequency behavior shows a prominent feature tied to the lattice phonons, which is not included in our present theory.
\begin{figure*}[!ht]
\begin{center}
\begin{minipage}{0.4\linewidth}    
\hspace{0.25cm} \vspace{0.2cm} \\
    (a) \vspace{0.1cm} \\
    \includegraphics[height=0.95\linewidth]{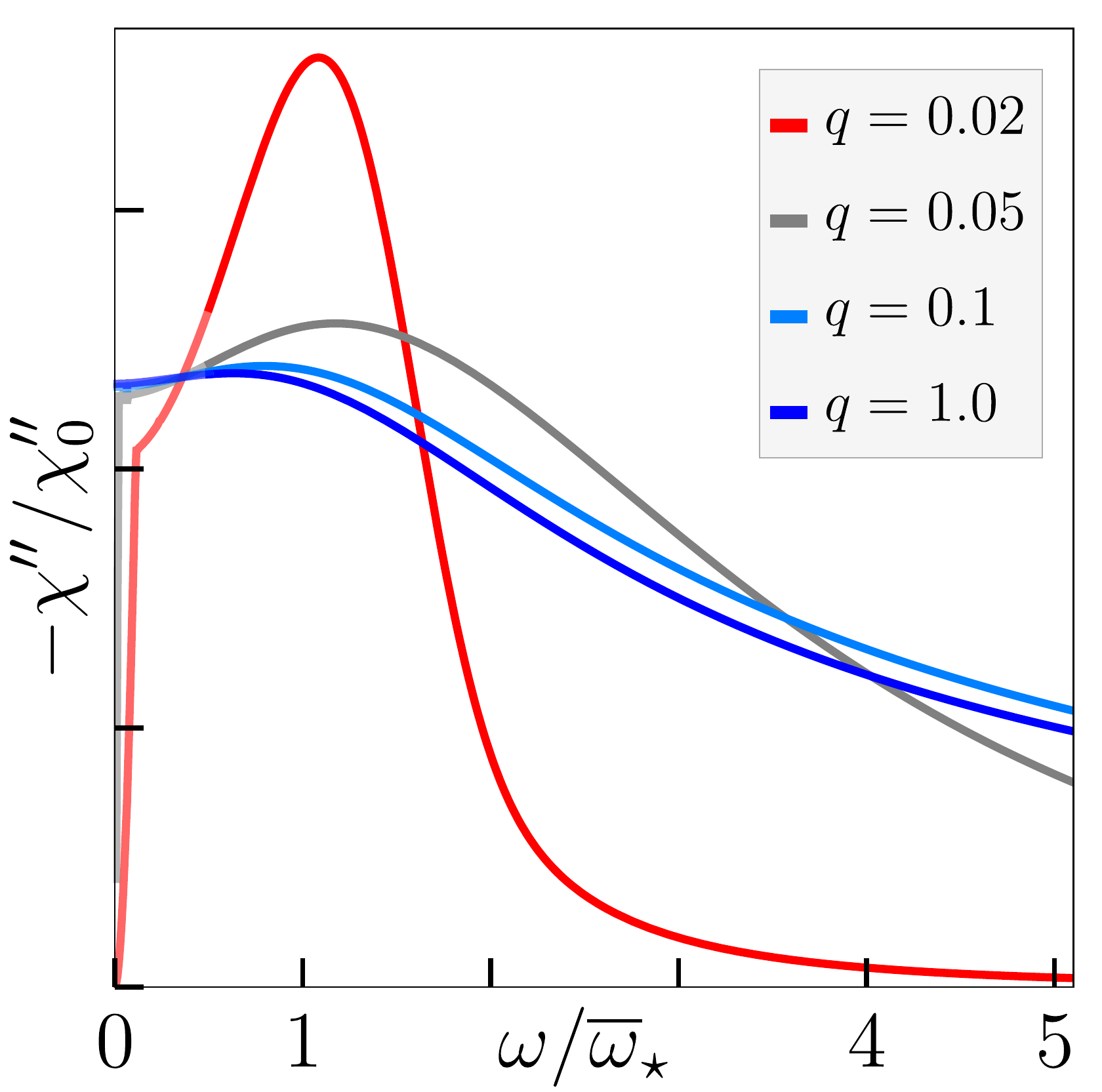}
\end{minipage}\hspace{0.03\linewidth}
\begin{minipage}{0.4\linewidth}
\hspace{0.25cm} \vspace{0.2cm} \\
    (b) \vspace{0.1cm} \\
    \includegraphics[height=0.95\linewidth]{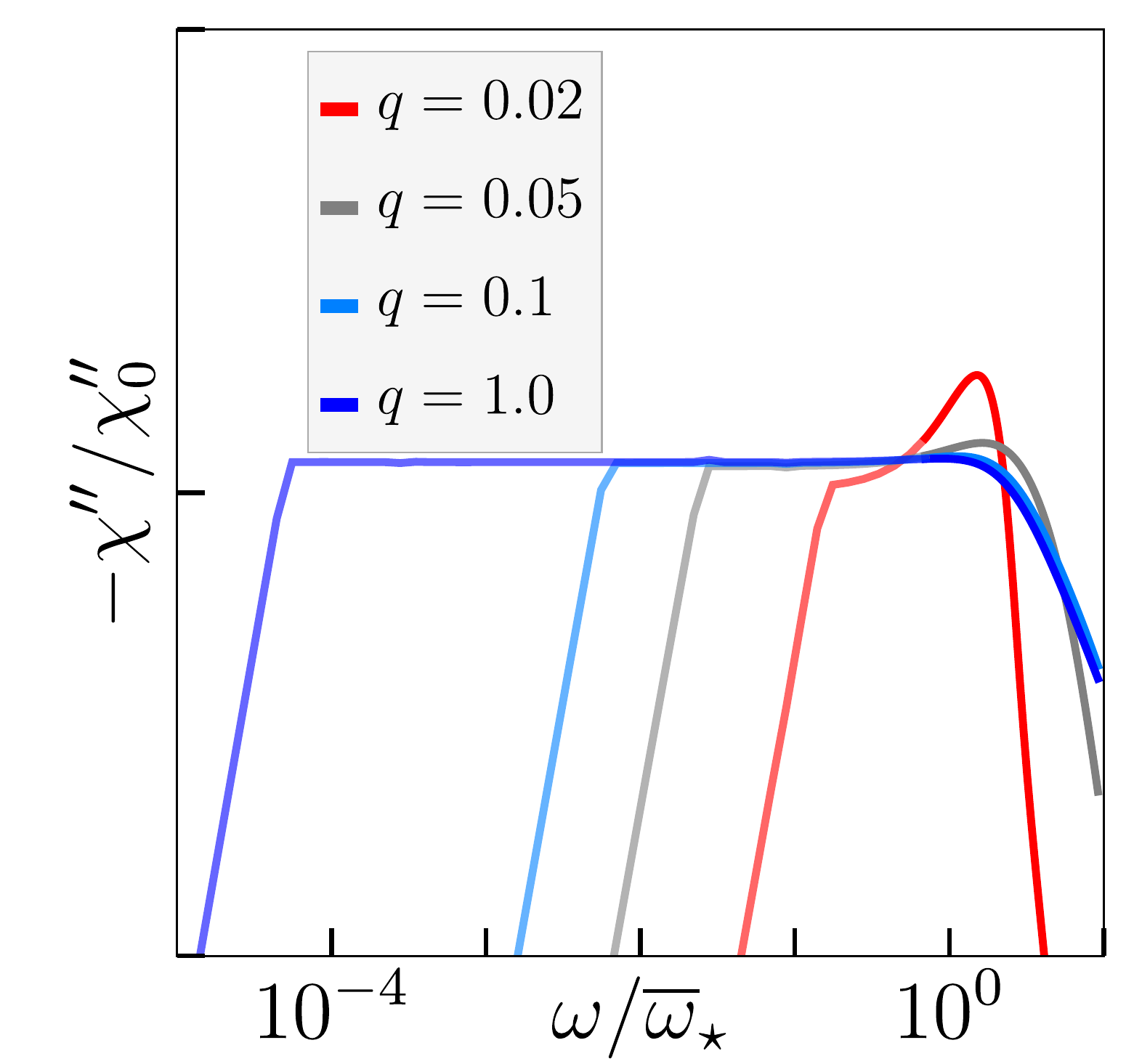}
\end{minipage}
\end{center}
\caption{Low frequency behavior of the disorder-averaged response, including an anomalous ``sharp'' feature at the lowest frequencies where the plateau crosses over into a power-law regime enforcing $\chi''(q,0)=0$. (a) is shown on a linear-linear scale, while (b) is shown on a log-log scale.
\label{fig:sharpfeat}}
\end{figure*}

\end{document}